%% file: sample-sigconf.tex
\documentclass[acmsmall]{acmart}

\AtBeginDocument{%
  \providecommand\BibTeX{{%
    \normalfont B\kern-0.5em{\scshape i\kern-0.25em b}\kern-0.8em\TeX}}}

\copyrightyear{2025}
\acmYear{2025}
\setcopyright{acmlicensed}

\acmDOI{XXXXXXX.XXXXXXX}
\acmJournal{JACM}

\usepackage{booktabs}
\usepackage{multirow}
\usepackage{algorithm}
\usepackage{algorithmic}
\usepackage{enumitem}
\usepackage{enumerate}
\usepackage{xcolor}
\usepackage{graphicx}
\usepackage{subcaption}
\usepackage{diagbox} 
\usepackage{amsmath, graphicx, hyperref}
\usepackage{geometry}
\usepackage{makecell}

\newcommand{\modelname}{ReLLaX}
\newcommand{\eg}{\emph{e.g.}}
\newcommand{\ie}{\emph{i.e.}}

\begin{document}

\title{Full-Stack Optimized Large Language Models for Lifelong Sequential Behavior Comprehension in Recommendation}

\author{Rong Shan}
\email{shanrong@sjtu.edu.cn}
\authornote{Equal contribution.}
\affiliation{
  \institution{Shanghai Jiao Tong University}
  \city{Shanghai}
  \country{China}
}

\author{Jiachen Zhu}
\email{gebro13@sjtu.edu.cn}
\authornotemark[1]
\affiliation{
  \institution{Shanghai Jiao Tong University}
  \city{Shanghai}
  \country{China}
}

\author{Jianghao Lin}
\email{chiangel@sjtu.edu.cn}
\affiliation{
  \institution{Shanghai Jiao Tong University}
  \city{Shanghai}
  \country{China}
}

\author{Chenxu Zhu}
\email{zhuchenxu1@huawei.com}
\affiliation{
  \institution{Huawei Noah's Ark Lab}
  \city{Shenzhen}
  \country{China}
}

\author{Bo Chen}
\email{chenbo116@huawei.com}
\affiliation{
  \institution{Huawei Noah's Ark Lab}
  \city{Shenzhen}
  \country{China}
}

\author{Ruiming Tang}
\email{tangruiming@huawei.com}
\affiliation{
  \institution{Huawei Noah's Ark Lab}
  \city{Shenzhen}
  \country{China}
}
\author{Yong Yu}
\email{yyu@sjtu.edu.cn}
\affiliation{
  \institution{Shanghai Jiao Tong University}
  \city{Shanghai}
  \country{China}
}
\author{Weinan Zhang}
\authornote{Weinan Zhang is the corresponding author.}
\email{wnzhang@sjtu.edu.cn}
\affiliation{
  \institution{Shanghai Jiao Tong University}
  \city{Shanghai}
  \country{China}
}

\renewcommand{\shortauthors}{Rong Shan et al.}

\begin{abstract}

As large language models (LLMs) achieves remarkable success in natural language processing (NLP) domains, LLM-enhanced recommender systems have received much attention and are being actively explored currently. 
In this paper, we focus on adapting and enhancing large language models for recommendation tasks.
First and foremost, we identify and formulate  \emph{the lifelong sequential behavior incomprehension problem} for LLMs in recommendation realms, \ie, LLMs fail to effectively extract useful information from a pure textual context of long user behavior sequence, even if the length of context is well below the context limitation of LLMs. 
To address such an issue and improve the recommendation performance of LLMs, we propose a novel framework, namely \underline{\textbf{R}}etrieval-\underline{\textbf{e}}nhanced \underline{\textbf{L}}arge \underline{\textbf{La}}nguage models \textbf{\underline{Plus}} (ReLLaX), which provides full-stack optimization from three perspectives, \ie, data, prompt and parameter. 
For data-level enhancement, we design semantic user behavior retrieval (SUBR) to reduce the heterogeneity of the behavior sequence, thus lowering the difficulty for LLMs to extract the essential information from user behavior sequences. Although SUBR can improve the data quality, further increase of the sequence length will still raise its heterogeneity to a level where LLMs can no longer comprehend it. Hence, we further propose to perform prompt-level and parameter-level enhancement, with the integration of conventional recommendation models (CRMs).
As for prompt-level enhancement, we apply soft prompt augmentation (SPA) to explicitly inject collaborative knowledge from  CRMs into the prompt. The item representations of LLMs are thus more aligned with recommendation, helping LLMs better explore the item relationships in the sequence and facilitating comprehension. Finally for parameter-level enhancement, we propose component fully-interactive LoRA (CFLoRA). By enabling sufficient interaction between the LoRA atom components, the expressive ability of LoRA is extended, making the parameters effectively capture more sequence information. Moreover, we present new perspectives to compare current LoRA-based LLM4Rec methods, \ie\ from both a composite and a decomposed view. We theoretically demonstrate that the ways they employ LoRA for recommendation are degraded versions of our CFLoRA, with different constraints on atom component interactions.
 Extensive experiments are conducted on three real-world public datasets to demonstrate the superiority of ReLLaX compared with existing baseline models, as well as its capability to alleviating lifelong sequential behavior incomprehension. 
Our code is available\footnote{ \url{https://github.com/LaVieEnRose365/ReLLaX}}.

\end{abstract}



\maketitle

\input{text/introduction}
\input{text/formulation}

\input{text/methodology}

\input{text/experiment}

\input{text/related_work}
\input{text/conclusion}


\bibliographystyle{ACM-Reference-Format}
\bibliography{acmart}

\input{text/appendix}

\end{document}

%% file: text/introduction.tex
\section{Introduction}

\begin{figure}[t]
  \centering
  \includegraphics[width=0.8\textwidth]{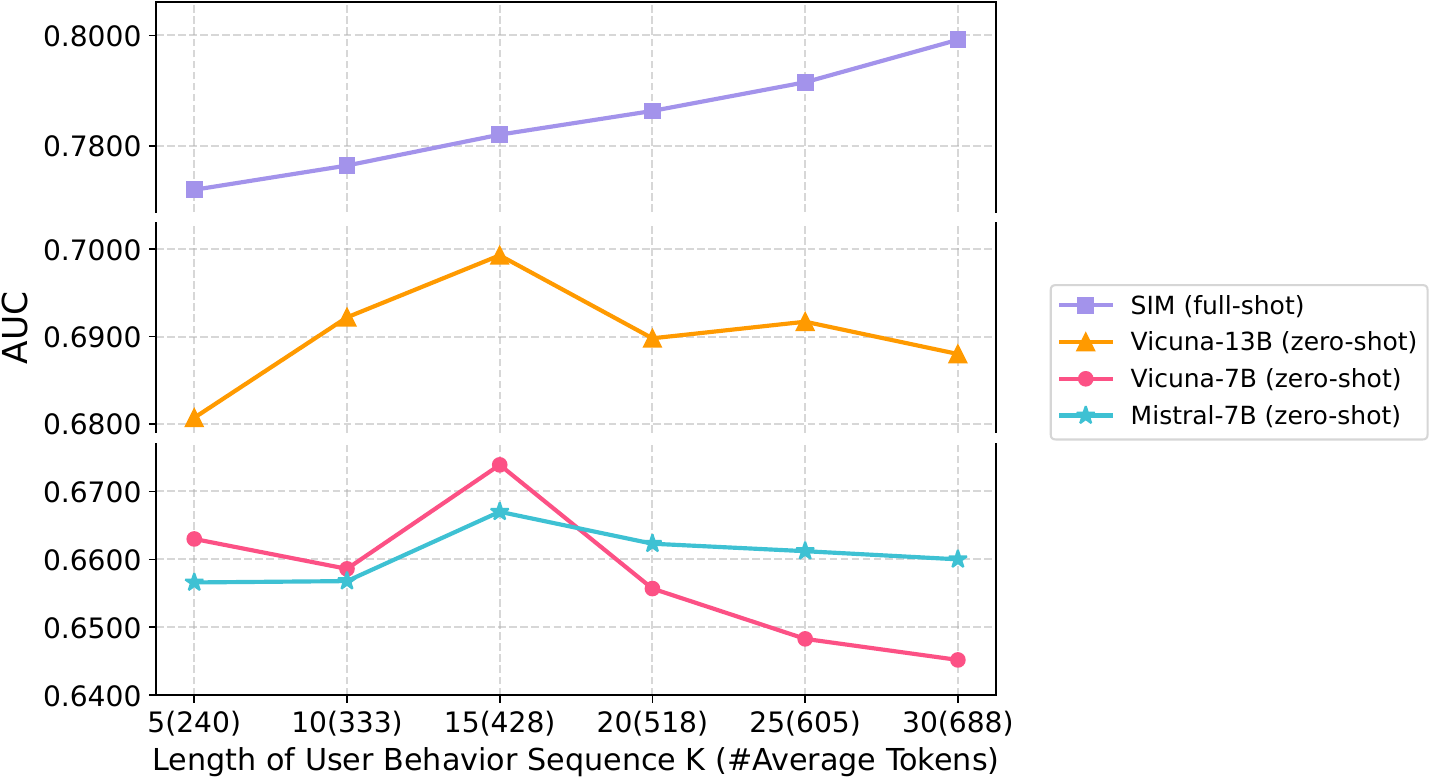}
  \caption{The illustration of lifelong sequential behavior incomprehension problem for LLMs. We report the AUC performance of SIM and different LLMs on MovieLens-1M dataset. 
  While SIM enjoys steady performance improvement as the length of behavior sequence $K$ grows, the LLMs
  generally only peaks at $K=15$ and fails to extract the useful information with further longer sequences (\ie, $K>15$).}
  \label{fig:illustration long context problem}
\end{figure}
Recommender systems play an important role in various online applications to address the challenge of information overload and fulfill users' information needs~\cite{DeepFM,xi2023bird,xi2023towards}. Meanwhile, large language models (LLMs) have achieved significant advancements in the natural language processing (NLP) field, demonstrating remarkable abilities to generate human-like text across diverse tasks~\cite{brown2020language,touvron2023llama,wang2023recmind,zhang2023memory}. 
As a result, recent studies have begun to investigate the integration of LLMs into recommender systems~\cite{lin2023can,hou2023large,bao2023tallrec,lin2024clickprompt}. These works directly leverage LLMs for various recommendation tasks, such as listwise ranking and pointwise scoring, revealing their promising performance in enhancing recommendation systems~\cite{zhang2023recommendation,bao2023tallrec}.

In this paper, we focus on adapting and enhancing large language models (LLMs) for recommendation tasks. Specifically, we identify and address the challenge of \textbf{lifelong sequential behavior incomprehension}, which refers to \emph{the inability of LLMs to effectively extract useful information from long user behavior sequences, even when the length of context is well below their context window limitations.} This issue is illustrated in Figure~\ref{fig:illustration long context problem}, where we evaluate LLMs of various types and sizes to highlight the universality of the problem.
As shown in the figure, traditional recommendation models (e.g., SIM) demonstrate steady performance improvements as the length of the user behavior sequence \(K\) increases. This indicates that longer sequences can indeed provide more useful information for recommendation tasks. However, the performance of different LLMs generally peaks at a sequence length of \(K=15\) and declines for longer sequences (\(K > 15\)), despite the total number of tokens being far below their context window limits (e.g., 2048 tokens for Vicuna-13B~\cite{vicuna2023} and 32,768 tokens for Mistral-7B~\cite{jiang2023mistral}). 

Interestingly, in typical NLP tasks, LLMs exhibit exceptional performance when processing similar context lengths (around 600+ tokens)~\cite{qin2024large, patil2024review}. This discrepancy suggests that the incomprehension of long user behavior sequences is a universal and domain-specific challenge for LLMs in recommendation systems. The task of inferring user preferences for a candidate item based on a given user profile and behavior history is inherently complex, requiring nuanced reasoning capabilities that current LLMs struggle to achieve in this context. 


To address the lifelong sequential behavior incomprehension problem, we propose a novel framework to develop \underline{\textbf{R}}etrieval-\underline{\textbf{e}}nhanced \underline{\textbf{L}}arge \underline{\textbf{La}}nguage models
\underline{\textbf{Plus}} (ReLLaX) for recommendation tasks, which provides full-stack optimization for the LLM4Rec paradigm. Specifically, we offer solutions from three perspectives: data, prompt, and parameter. 

From the data perspective, to address the performance issues associated with long sequences, we propose \textbf{Semantic User Behavior Retrieval (SUBR)}. Instead of simply truncating the top-$K$ most recent behaviors, SUBR selects the top-$K$ semantically relevant behaviors to the target item. This approach enhances the quality of data samples and lowers the heterogeneity of the user behavior sequence, reducing the difficulty for LLMs in extracting meaningful information from user behavior sequences. 

Nevertheless, further increase of sequence length will still raise the heterogeneity to a high level, where the LLMs can no longer understand it. Therefore, we additionally propose prompt-level and parameter-level techniques to enhance LLMs' comprehension of the behavior sequence.

From the prompt dimension, instead of relying solely on hard prompts composed of pure textual tokens, we propose a \textbf{Soft Prompt Augmentation (SPA)} module and construct more comprehensive prompts. Specifically, for each item, we employ a lightweight projector to transform its ID representations obtained from a pretrained sequential recommendation model into soft prompt tokens. These soft prompt tokens are then concatenated with the original textual tokens in the hard prompt. In this way,  SPA enhances the vanilla item representations in language space of LLMs and makes them more aligned with recommendation. This helps LLMs better explore the item relationships in the sequence, thus improving their ability to handle the long behavior sequences. 

\begin{figure}[t]
  \centering
  
  \includegraphics[width=\textwidth]{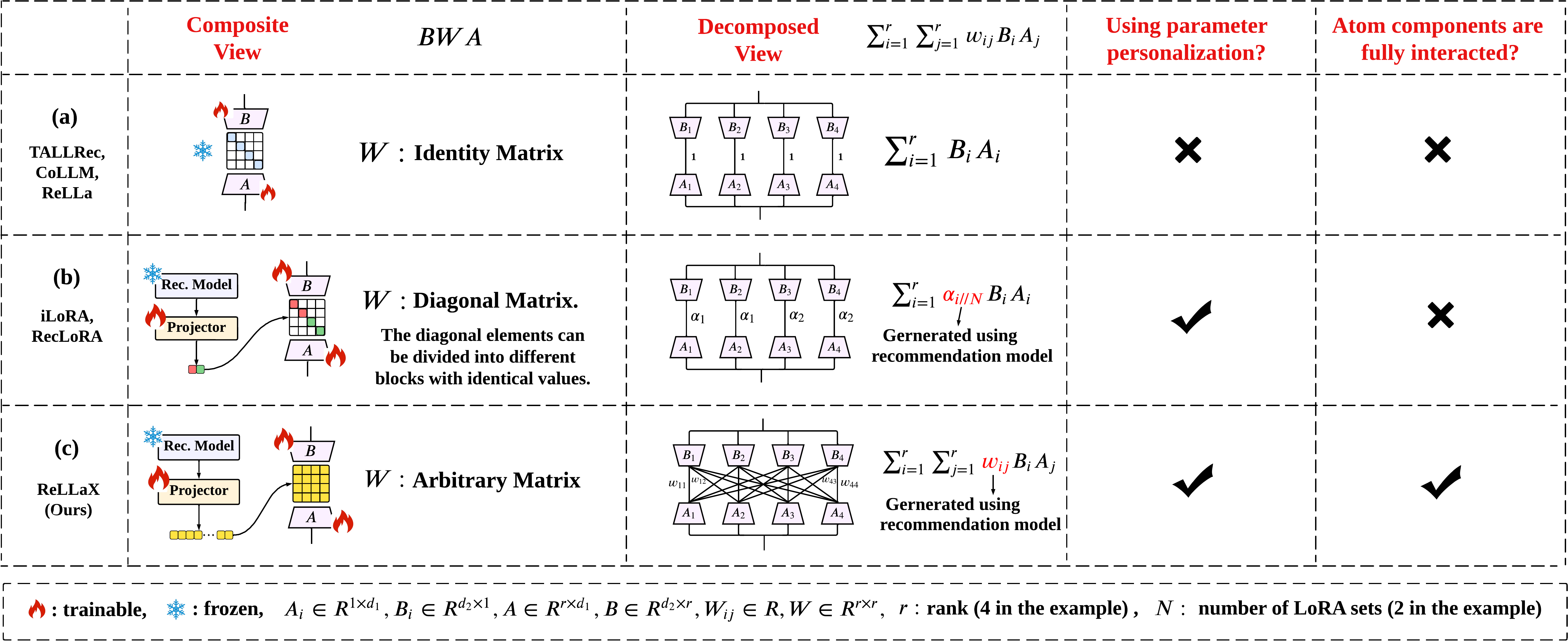}
  \caption{Comprehensive comparison of the ways different LLM4Rec methods employ LoRA for finetuning.}

  \label{fig:lora discussion}
\end{figure}

From the parameter perspective, current LLM4Rec methods~\cite{bao2023tallrec, lin2024rella, zhu2024lifelong, kong2024customizing} usually leverages low-rank adaptation (LoRA)~\cite{hu2021lora} for parameter-efficient finetuning (PEFT). However, due to the issue of lifelong sequential behavior incomprehension, vanilla LoRA parameters may not expressive enough to effectively capture information from the long sequence. To address this, we propose component fully-interactive LoRA (CFLoRA) module, where we decompose LoRA matrices into atom components (\ie, vectors) and enable full interaction between them. Inspired by previous works on LoRA personalization~\cite{zhu2024lifelong, kong2024customizing}, we guide the interaction strength  with the integration of conventional recommendation models. Through sufficient interaction of atom components, the LoRA parameters are more expressive and can  capture more long-sequence information effectively, facilitating LLMs' comprehension ability. Moreover, as a novel LoRA-based method, we also compare our CFLoRA module with the ways existing LLM4Rec methods~\cite{zhu2024lifelong, kong2024customizing, lin2024rella, bao2023tallrec, zhang2023collm, liao2024llara} employ LoRA for recommendation. The summarization of our analysis is shown in Figure~\ref{fig:lora discussion}. Specifically, we present new perspectives to analyze LoRA parameters, \ie, from both a composite and a decomposed view. We theoretically illustrate that these LoRA-based LLM4Rec methods are merely special cases of the CFLoRA module, and CFLoRA can degrade into them with different constraints on the atom components interactions. CFLoRA not only allows full interaction between LoRA atom components under the decomposed view, but also is elegant and  
easy to implement under the composite view. In comparison, the  interactions among atom components in existing methods remain partial and not sufficient.

Main contributions of this paper are in three folds:
\begin{itemize}
    \item To the best of our knowledge, we are the first to identify and formally define the \textbf{lifelong sequential behavior incomprehension} problem in LLM-based recommendation systems, \eg\  LLMs struggle to comprehend textual contexts containing long user behavior sequences, even when the context length is well within the model's context window limitation.
    \item We propose a comprehensive framework, \text{\modelname}, which implements full-stack optimization for long-sequence challenges from three perspectives: data, prompt, and parameter. Specifically, we introduce a retrieval-based (\ie, SUBR) method to enhance the quality of data samples, soft prompt augmentation (SPA) to construct more comprehensive prompts and design a novel component Fully-interactive LoRA (CFLoRA) module for instruction tuning.
    
    \item We also conduct a theoretical and comprehensive analysis to compare the CFLoRA module with other existing LoRA-based LLM4Rec methods under both a composite and a decomposed view. In a decomposed form, CFLoRA ensures more sufficient interaction among LoRA atom components, leading to improved performance. Besides, CFLoRA maintains a concise and intuitive composite form that can be elegantly and efficiently implemented via matrix multiplication.
    \item Extensive experiments on three real-world public datasets validate the effectiveness of our method compared with existing baselines. 

\end{itemize}

This paper is a substantial extension of its conference version~\cite{lin2024rella}. The major differences between the journal version and the conference paper are:
\begin{itemize}
    \item In this paper, we propose \modelname\ as a more powerful framework against the original ReLLa proposed in the conference paper. Apart from leveraging retrieval to improve the hard prompt, we design soft prompt augmentation (SPA) to further reduce the difficulty of user sequence modeling for LLMs and inject collaborative knowledge. Moreover, we design component fully-interactive LoRA (CFLoRA) to enhance the expressive ability of LoRA in recommendation.
    \item We revisit the ways how ReLLa and other LoRA-based LLM4Rec methods employ LoRA from both a composite and decomposed view, and theoretically illustrate the connections and difference between our proposed CFLoRA module and these works. We show that the ways these methods employ LoRA are actually degraded versions of CFLoRA with different constraints on atom components interaction. In comparison, CFLoRA in \modelname\ leads to more interactions and fusions between atom components, with the integration of user lifelong sequence information.
    \item We provide the experimental results compared with several recent LLM4Rec methods that are proposed later than our conference paper.

\end{itemize}

%% file: text/formulation.tex
\section{Preliminaries}
\label{sec:preliminary}
In this paper, we focus on click-through rate (CTR) prediction, a critical component of recommender systems that estimates the probability of a user clicking on a target item within a given context. The training dataset for CTR prediction can be represented as $\mathcal{D}=\{(x_i, y_i)\}_{i=1}^N$, where $N$ denotes the number of data samples. In next parts, we give the formulation of CTR prediction tasks for conventional recommendation models and introduce how to leverage LLMs for pointwise scoring, as well as the low-rank adaptation (LoRA) for instruction tuning.

\subsection{CTR Problem Formulation}
Click-through rate (CTR) prediction is a fundamental task in recommendation systems, aimed at estimating the probability of a user clicking on a specific item. Formally, let $\mathcal{U}$ denote the set of users with their profiles (\eg\ age or gender) and $\mathcal{I}$ denote the set of items with their attributes (\eg\ category or brand). Each user $u \in \mathcal{U}$ has a has a chronological interaction sequence $H_u = [i_l]_{l=1}^L$, where $i_l \in \mathcal{I}$ is the $l$-th item interacted by the user $u$ and $L$ is the length of the interaction sequence. Given a target item $i_c$, the objective of is to predict the probability $\hat{y}$ of the user $u$ clicking on $i_c$, which can be represented as:
\begin{equation}
\hat{y} = P(y = 1 \mid u, H_u, i_c),
\end{equation}
where $y = 1$ indicates a click. Let $x = (u,H_u, i_c) $ denote the current input, and it can be expressed in two modalities: (1) ID modality $x^{ID}$, generated using a one-hot encoder, which would be taken as input for conventional recommendation models, and (2) text modality $x^{text}$, which will be directly fed into the LLMs.

\subsection{Conventional Recommendation Models}
\label{sec:CRM}
Typically, conventional recommendation models take $x^{ID}$ as inputs, and then encode $x^{ID}$ into dense representations with the embedding layers. Let $r_u, e_{i_c}\in \mathbb{R}^{d_e}$ denote the representation for the user $u$ and target item $i_c$ respectively, and $d_e$ is the embedding size. $E_u^{ID}=[e_{i_l}]_{l=1}^L\in \mathbb{R}^{L\times d_e}$  denotes the representations of items in the interaction sequence. Next, different functions $f(\cdot)$ can be leveraged to aggregate the dense representations for feature interaction or user sequence modeling, the process can be written as:
\begin{equation}
\label{eq: getting CRM representation}
h = f(r_u, E_u, e_{i_c}) \in \mathbb{R}^{d_h},
\end{equation}
where $h$ is the final representation, and $d_h$ is the hidden dimension. Finally, a classification head $g$, which is often a linear layer with sigmoid function, maps the final representation into the predicted click-through rate (CTR) $\hat{y}$:
\begin{equation}
\hat{y} = g(h) \in[0,1].
\end{equation}

The model can be trained in an end-to-end manner with the binary cross-entropy (BCE) loss:
\begin{equation}
\mathcal{L} = -\frac{1}{|\mathcal{D}|} \sum\nolimits_{j=1}^{|\mathcal{D}|}[y_j\operatorname{log}\hat{y}_j+(1-y_j)\operatorname{log}(1-\hat{y}_j)].
\end{equation}

    

\subsection{Pointwise Scoring with Large Language Models (LLMs)}
\label{sec:pointwise scoring}

Let $\mathcal{M}_\Theta$ denote a large language model with parameters $\Theta$. The LLM takes the discrete tokens of $x^{text}$ as input, and generates the next token $\hat{y}_i^{text}$ as the output, the process of which can be formulated as follows:
\begin{equation}
\label{eq: llm predict token}
\begin{aligned}
    s &= \mathcal{M}_\Theta(x^{text})\,\in\mathbb{R}^{V}, \\
    p &= \operatorname{Softmax}(s)\,\in\mathbb{R}^V,\\
    \hat{y}^{text} &\sim p \,,
\end{aligned}
\end{equation}
where $V$ is the vocabulary size, and $\hat{y}^{text}$ is the next predicted token sampled from the probability distribution $p$. 

Nevertheless, CTR prediction demands the model to conduct pointwise scoring, in which case the output should be a floating-point number $\hat{y}\in[0,1]$ instead of a discrete token $\hat{y}^{text}$.
Therefore, we follow previous works~\cite{bao2023tallrec,zhang2023prompt} and intercept the estimated scores $s\in\mathbb{R}^V$ of the binary key answer words ``Yes'' and ``No''. Then we conduct a bidimensional softmax over them to predict the CTR. Specifically, suppose the token indices for ``Yes'' and ``No'' in the LLM vocabulary are $m$ and $n$ respectively,
the pointwise scoring of LLMs for CTR prediction can be written as:
\begin{equation}
\begin{aligned}
    \hat{y} = \frac{\exp(s_{m})}{\exp(s_{m})+\exp(s_{n})} \,\in[0,1].
\end{aligned}
\end{equation}

It is noteworthy that such an estimated click-through rate $\hat{y}$ is only used on the testing set for evaluation, which also enables zero-shot inference. For training, the common paradigm of instruction tuning is preserved, which leverages the following causal language modeling objective:
\begin{equation}
\label{eq: causal LM}
\max_{\Theta}\sum\nolimits_{(x^{text},y^{text})\in\mathcal{D}}\, \sum\nolimits_{j=1}^{|y^{text}|}\log P_{\Theta}(y_{j}^{text}|x^{text},y^{text}_{<j}),
\end{equation}
where $y_j$ is the $j$-th token of the textual output $y$, and $y_{<j}$ denotes the tokens before $y_j$.

\subsection{LoRA for Efficient Adaptation}
\label{sec:lora}
Full finetuning of LLMs is time-consuming and resource-intensive, and low-rank adaptation (LoRA) serves as an efficient alternative. It introduces trainable low-rank matrices to approximate the updates of pretrained weights, while the pretrained weights are frozen. Specifically, LoRA employs a low-rank decomposition on the update of the weights $\Delta \Theta$. The process can be represented as: \begin{equation}
\Delta \Theta = B A,
\end{equation}
where $A \in \mathbb{R}^{r\times d_{down}}, B \in \mathbb{R}^{d_{up}\times r}$ are down-projection and up-projection matrices respectively. $A$ can be randomly initialized, while  $B$ is often initialized as zero matrix to ensure the output equal to the original value. The rank $r$ is much smaller than both $d_{down}$ and $d_{up}$, improving the efficiency of model adaptation. By applying LoRA on the LLM transformer layers, the causal language modeling objective in Equation~\ref{eq: causal LM} can be rewritten as:
\begin{equation}
\label{eq: causal LM}
\max_{\Delta\Theta }\sum\nolimits_{(x^{text},y^{text})\in\mathcal{D}}\, \sum\nolimits_{j=1}^{|y^{text}|}\log P_{\Theta+\Delta\Theta}(y_{j}^{text}|x^{text},y_{<j}^{text}),
\end{equation}
where $\Theta$ is fixed and only $\Delta \Theta$ is trainable during the instruction tuning process. Correspondingly, the next token logits $s$ in Equation~\ref{eq: llm predict token} when evaluation on the testing set can be reformulated as:
\begin{equation}
\label{eq: lora infer}
\begin{aligned}
    s &= \mathcal{M}_{\Theta+\Delta\Theta}(x^{text}).
\end{aligned}
\end{equation}




%% file: text/methodology.tex
\section{Methodology}
\label{sec:method}

\begin{figure}[t]
  \centering
  
  \includegraphics[width=\textwidth]{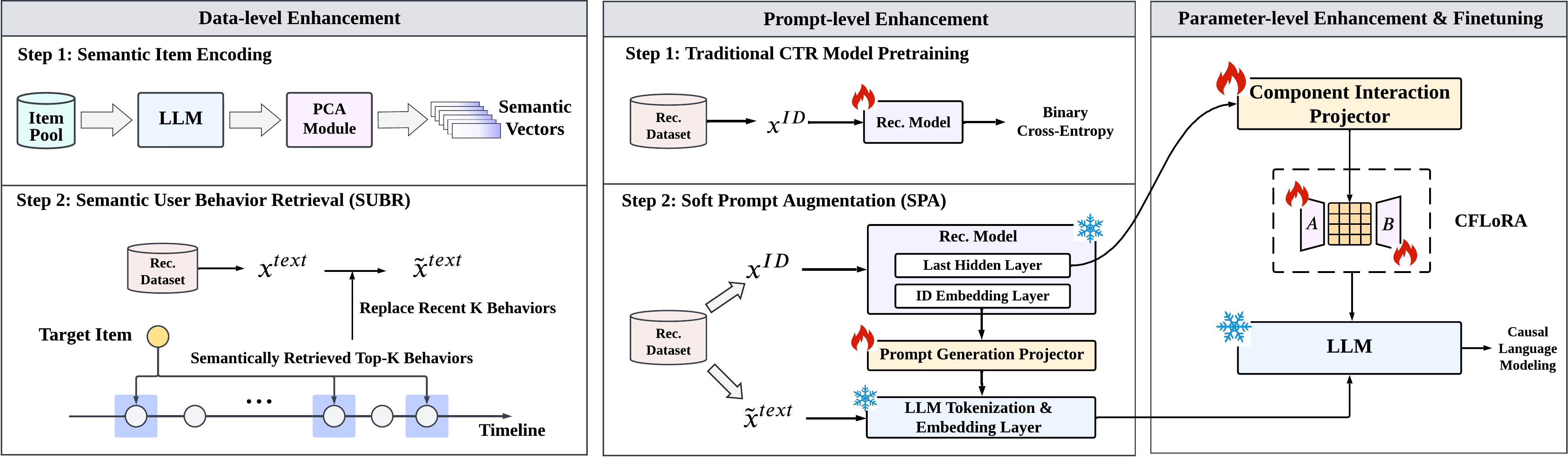}
  \caption{The overall framework of our proposed ReLLaX, which proposes a full-stack optimization for LLMs to address the lifelong sequence incomprehension in recommendation. The enhancement lies in three different levels, \eg\ data, prompt and parameter.}

  \label{fig:rellax_framework}
\end{figure}

  


In this section, we introduce our proposed ReLLaX framework in details. First, we will provide an overview of the whole framework. Then we elaborate on our key techniques in three aspects, \eg\ data, prompt and parameter. 

\subsection{Overview of ReLLaX}
As illustrated in Figure~\ref{fig:rellax_framework}, ReLLaX proposes a full-stack optimization framework for LLMs to tackle the problem of lifelong sequence incomprehension in recommendation, where our developed techniques provide (1) data-level, (2) prompt-level and (3) parameter-level enhancement, respectively. 

\textbf{Data-level Enhancement.} To improve the data quality, we propose to perform semantic user behavior retrieval (SUBR) over the textual data samples. We first obtain the semantic vectors for each item utilizing the large language model, based on which we can retrieve the top-$K$ \textit{semantically relevant} behaviors for each textual data sample $x^{text}$. Then these retrieved behaviors can substitute the original top-$K$ \textit{recent} behaviors, resulting in the retrieval-enhanced data sample $\tilde{x}^{text}$. The heterogeneity of items in the sequence is lowered, reducing the difficulty for LLMs to extract useful information from the sequence. Note that SUBR serves for data augmentation purpose, and $\tilde{x}^{text}$ will be the input for LLMs to reason over, rather than $x^{text}$.

\textbf{Prompt-level Enhancement.}  We propose to conduct soft prompt augmentation (SPA) to make the vanilla item representations of LLMs more aligned with recommendation. Firstly, we pretrain a conventional recommendation model (CRM), whose item ID embeddings lie in recommendation space and encode collaborative knowledge well. Secondly, to adapt these ID embeddings into the language space of LLM, we leverage a lightweight projector to convert them into soft prompt tokens, which are then concatenated with the original textual tokens of the items. This comprehensive prompting design helps LLMs model the item relationships in the sequence better, thus improving LLMs' ability to handle inputs with long behavior sequences.

\textbf{Parameter-level Enhancement.} We develop component fully-interactive LoRA (CFLoRA) to construct personalized parameters more effectively for efficient finetuning. We first harness the pretrained CRM to obtain the final representation for each data sample, which encodes all collaborative information for recommendation. Based on the these representations, component interaction matrices can be generated using a lightweight projector, which would be inserted between the vanilla LoRA decomposition matrices. The process operates in a composite view which focuses on LoRA matrices. Notably, in the corresponding decomposed view where we investigate the LoRA atom components, CFLoRA allows more sufficient interaction between LoRA atom components than existing LoRA-based LLM4Rec methods. This extends the expressive capability of LoRA, and makes the LoRA parameters effectively capture more sequence information.

\subsection{Semantic User Behavior Retrieval}
\label{sec: SUBR}
The user behavior sequences often consist of diverse, heterogeneous, and sometimes random or noisy item clicks~\cite{wang2021denoising, wang2023efficient, lin2024rella}, making it difficult for LLMs to extract useful information from the sequences. Hence, we propose semantic user behavior retrieval (SUBR) to improve the data quality and  homogenize the sequence. SUBR operates by substituting the simply truncated most recent $K$ behaviors with the most semantically relevant $K$ behaviors towards the target item. The retrieved user behaviors serve to denoise the user behavior sequence and convey clearer and more essential user interests for the target item, while the original length of user behavior sequence is preserved.

\begin{figure}[h]
  \centering
  \includegraphics[width=0.8\textwidth]{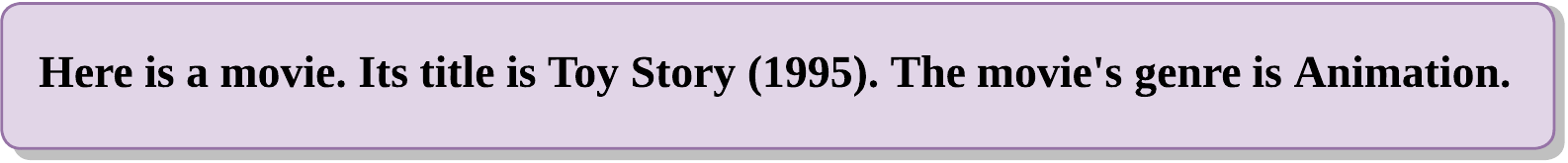}
  \caption{Illustration of descriptive text for an item (movie).}
  \label{fig:item description}
\end{figure}

Firstly, we perform \textit{semantic item encoding} to 
generate a semantic vector for each item. For the  $t$-th item in the pool, we construct a descriptive text through a hard prompt template (an example is shown in Figure~\ref{fig:item description}), and feed the text into the LLM. The hidden states from the last layer of the LLM are then averaged to produce a vector $z_t \in \mathbb{R}^{d_z}$, where $d_z$ is the hidden size of LLM (e.g., 4096 for Vicuna-7B and 5120 for Vicuna-13B). We further apply principal component analysis (PCA)~\cite{shlens2014tutorial} for both dimension reduction and denoising purposes, obtaining the final semantic vector $q_t \in \mathbb{R}^{d_q}$, where we set the dimension $d_q$ to 512. Now we can assess the semantic relevance between each pair of items by calculating the cosine similarity between their respective semantic vectors.

Next, we can employ \textit{semantic user behavior retrieval} on each sample $x^{text}$ to obtain the retrieval-enhanced counterpart $\tilde{x}^{text}$, which will be the new textual input for the LLM. Specifically, the original truncated top-$K$ recent behaviors are replaced with the top-$K$ semantically relevant behaviors of the target item, while maintaining a similar length of textual input. In this way, SUBR improves the data quality, reducing the difficulty for LLMs to comprehend the long user behavior sequences.

Moreover, We conduct an empirical study on MovieLens-1M dataset to
demonstrate the homogenizing effect of SUBR. Here, we define the heterogeneity score as the number of unique movie genres in a sequence of interacted items (i.e., movies). We illustrate two examples as follows:
\begin{itemize}
    \item $[\text{Comedy, Fiction, Comedy, Family}]$ $\rightarrow$ heterogeneity score=3
    \item $[\text{Fiction, Fiction , Child, Fiction}]$ $\rightarrow$ heterogeneity score=2
\end{itemize}

In Table~\ref{tab:heter score}, we report the averaged heterogeneity score of two sequence types w.r.t. different length $K$: (1) the sequence constructed by most recent behaviors. (2) the sequence consisting of most relevant behaviors generated by our proposed SUBR. We can observe that SUBR can lower the heterogeneity of the sequence, thus alleviating the incomprehension problem. However, as the heterogeneity score gradually increases when
the length $K$ grows, further increase of the sequence length will still raise the heterogeneity to a high level, where LLMs fail to understand it any more. For example, the sequence of top 60 relevant items have a similar heterogeneity of the sequence of top 30 recent items, where the heterogeneity is too high to make LLMs extract useful information. Therefore, in next parts, we further propose prompt-level and parameter-level enhancement of LLMs' sequence comprehension capability.

\begin{table}[t]
    \caption{The averaged heterogeneity scores of two sequence types w.r.t. different textual sequence length $K$.
    }
    \label{tab:heter score}
    \resizebox{0.8\textwidth}{!}{
    \renewcommand\arraystretch{1.1}
    \begin{tabular}{c|cccccccccc}
    \toprule
    \hline
    \multicolumn{1}{c|}{\multirow{2}{*}{Seq. Type}} & \multicolumn{9}{c}{MovieLens-1M} \\ 
    \multicolumn{1}{c|}{} & K=5 & K=10 & K=15 & K=20 & K=25 & K=30 & K=40 & K=50 & K=60 \\ 
   \hline 
   Top Recent (Origin) & 2.91 & 4.19 & 5.09 & 5.80 & 6.39 & 6.90 & 7.76 & 8.44 & 9.01 \\ 
   Top Relevant (Retrieval) & 2.44 & 3.37 & 4.01 & 4.51 & 4.94 & 5.32 & 5.78 & 6.43 & 6.96 \\
   \hline  
   \bottomrule          
\end{tabular}
}
\end{table}

\subsection{Soft Prompt Augmentation}
\label{sec: SPA}
Collaborative knowledge from conventional recommendation models (CRMs) can help LLMs capture user behavior patterns easier and improve recommendation performance, which has been revealed by previous works~\cite{zhang2023collm, liao2024llara}. Therefore, rather than relying solely on hard prompts composed of textual tokens, we employ soft prompt augmentation (SPA) to construct a more comprehensive prompt. Integrated with collaborative knowledge from CRMs, the comprehensive prompt can make the item representations in the language space more aligned with recommendation, and helps LLMs model the item relationships in the sequence better, which facilitates LLMs' understanding of the behavior sequences. Next, we dive into the details of the SPA module.

As described in Section~\ref{sec:preliminary}, each data sample $x$ can be expressed in two modalities, \eg\ ID modality $x^{ID}$ and text modality $x^{text}$ (now we use $\tilde{x}^{text}$ after SUBR). We first pretrain a conventional recommendation model with the full training set. Then for each data sample in ID modality $x^{ID}$, we can get the historical and target item embeddings $E^{ID}_{u,i_c}=[e_{i_1},e_{i_2}, \cdots, e_{i_L}, e_{i_c}]$ via looking up the pretrained CRM embedding tables. For each data sample in text modality $\tilde{x}^{text}$, we utilize the LLM tokenizer to tokenize it into $T=[t_l]_{l=1}^P$, where $t_l$ denotes $l$-th text token and $P$ denotes the number of tokens. These text tokens are further encoded into embeddings $E^{text}=[v_l]_{l=1}^P$, where $v_l\in \mathbb{R}^{d_v}$ is the token embedding of $t_l$ in the LLM embedding table, and $d_v$ is the LLM token embedding size.

Since the item ID embeddings $E^{ID}_{u,i_c}$ are pretrained in the recommendation space, not the language space where the LLM operates, we need to bridge the modality gap and adapt $E^{ID}_{u,i_c}$ into the language space. Specifically, we leverage a lightweight soft prompt generation projector, \eg\ a two-layer multilayer perceptron (MLP), to map the ID embeddings into soft prompt tokens $\hat{E}_{u,i_c} = [v_{i_1},v_{i_2}, \cdots, v_{i_L}, v_{i_c}]$. These soft prompt tokens are then appended behind their corresponding text token embeddings and result in the enhanced token embeddings $\tilde{E}$, which will be the final input fed to the LLM in our ReLLaX framework:
\begin{equation}
\begin{aligned}
    \tilde{E} = [v_1, \cdots, v_{n_1}, v_{i_1}, \cdots, v_{n_2}, v_{i_2}, \cdots, v_{n_L}, v_{i_L}, v_{n_{L+1}}, v_{i_c}],
\end{aligned}
\end{equation}
where $\{n_j\}_{j=1}^{L+1}$ are the last text token positions for the historical items and the target item.

\begin{figure}[t]
  \centering
  \includegraphics[width=0.8\textwidth]{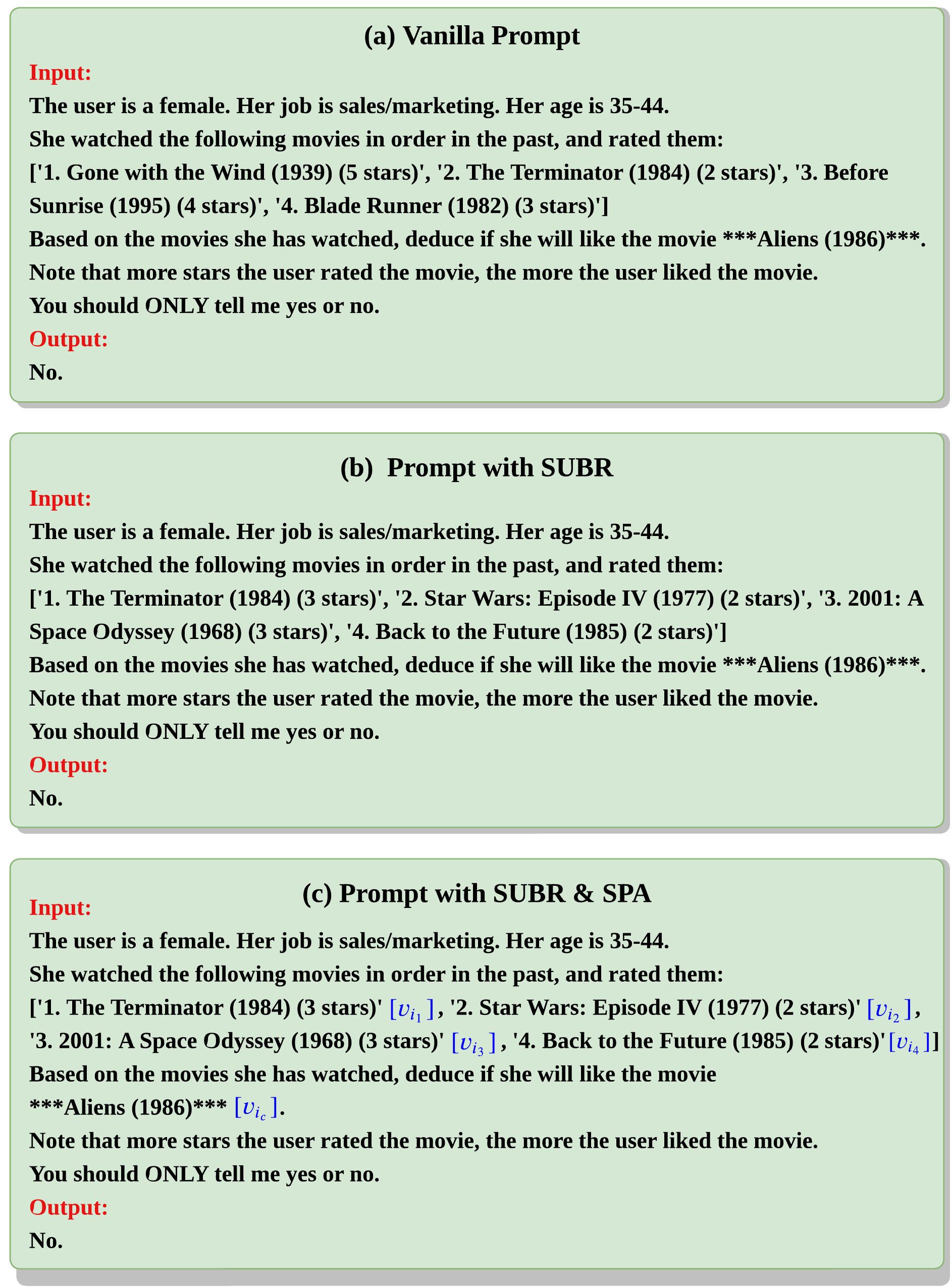}
  \caption{Examples of prompt templates for the MovieLens-1M dataset. \textbf{(a)} Vanilla prompt, which directly converts the recommendation data into text. \textbf{(b)} Hard prompt with SUBR, which replaces the most recent $K$ items in (a) with the most semantically relevant $K$ items towards the target item. \textbf{(c)} Comprehensive prompt with SUBR and SPA, extending (b) with the blue tokens, which represent the soft prompt tokens generated by SPA.}
  \label{fig:prompt comparison}
\end{figure}

As shown in Figure~\ref{fig:prompt comparison}, we provide several examples to further illustrate the vanilla prompt, the hard prompt with SUBR, as well as the comprehensive prompt with both SUBR and SPA. Our comprehensive prompting approach provides a more informative representation of items, combining the LLM's semantic information with the pretrained CRM model's collaborative filtering features. This facilitates the LLM's understanding of user history and improves its ability to perform CTR tasks.



\subsection{Component Fully-Interactive LoRA}
\label{sec:parameter-level enhancement}
Current LLM4rec methods~\cite{lin2024rella, bao2023tallrec, zhu2024lifelong, kong2024customizing, zhang2023collm} mainly leverage LoRA~\cite{hu2021lora} for parameter-efficient finetuning. However, due to the incomprehension problem, the vanilla LoRA parameters may not be expressive enough to effectively capture the long-sequence information. In this part, we first conduct a theoretical analysis of these methods by revisiting LoRA parameters, in both a \textbf{decomposed} and a \textbf{composite} view. Building on the analysis, we observe that the interaction between LoRA atom components is partial and not sufficient in these methods, leading to suboptimal effectiveness. Inspired by this observation, we propose component fully-interactive LoRA (CFLoRA), which allows more sufficient interaction between decomposed atom components, while maintaining a concise and elegant composite form. Through fully interaction of atom components, CFLoRA parameters can effectively capture more information from the user behavior sequence, promoting LLMs' ability of understanding the sequence. The comparison of our method with others is summarized in Figure~\ref{fig:lora discussion}.

\subsubsection{Decomposed View} 
First, we examine the LoRA matrices from a decomposed view, where we focus on the vectors composing the matrices. As stated in Section~\ref{sec:lora}, the vanilla LoRA approximates the update of weights by the multiplication of the up-projection matrix $B \in \mathbb{R}^{d_{up} \times r}$ and the down-projection matrix $A \in \mathbb{R}^{r\times d_{down}}$.  If we partition $A$ and $B$ along the $r$-dimension into vectors, we can get:
\begin{equation}
    B = [B_1, B_2, \cdots, B_r], A^T = [A_1^T,A_2^T, \cdots, A_r^T],
\end{equation}
where $B_j \in \mathbb{R}^{d_{down} \times 1}$ is the $j$-th column vector for $B$, and  $A_j \in \mathbb{R}^{1\times d_{up}}$ is the $j$-th row vector for $A$. We regard these vectors as atom components of the LoRA matrices, as they cannot be decomposed further. With these atom components, the vanilla LoRA matrices can be expressed as:
\begin{equation}
\label{eq: lora 1}
\begin{aligned}
    \Delta\Theta_1 =BA = \sum_{j=1}^r B_jA_j.   
\end{aligned}
\end{equation}
Hence, the vanilla LoRA matrices can be viewed as the aggregation of one-to-one interaction between $B_j$ and $A_j$, where the aggregation weight is fixed to 1.

Although vanilla LoRA has demonstrated promising performance in many LLM4Rec works~\cite{bao2023tallrec, zhang2023collm, liao2024llara, lin2024rella}, some recent works~\cite{zhu2024lifelong, kong2024customizing} further propose to construct personalized LoRA parameters. They generally prepare mutiple sets of LoRA parameters, and use conventional recommendation models to generate the scaling aggregation weights for different sets of LoRA. Therefore, following the original papers~\cite{zhu2024lifelong, kong2024customizing}, the update can be formulated as:
\begin{equation}
\label{eq: ilora 1}
\begin{aligned}
    \Delta\Theta_2 =\sum_{j=1}^N\alpha_j \hat{B}_j\hat{A}_j,   
\end{aligned}
\end{equation}
where $N$ denotes the number of LoRA sets. $\hat{B}_j \in \mathbb{R}^{d_{up}\times r_1}, \hat{A}_j\in \mathbb{R}^{r_1\times d_{down}}$ are $j$-th set of LoRA matrices, and $\alpha_j$ is the scaling weight generated by CRM for them. $r_1$ is their rank. In our decomposed view, we can rewrite Equation~\ref{eq: ilora 1} into:
\begin{equation}
\label{eq: ilora 2}
\begin{aligned}
    \Delta\Theta_2 =\sum_{j=1}^N \sum_{k=1}^{r_1}\alpha_jB^j_kA^j_k,
\end{aligned}
\end{equation}
where $B^j_k$ and $A^j_k$ are $k$-th atom components of $j$-th LoRA matrices. Furthermore, to make Equation~\ref{eq: ilora 2} more concise, we can absorb the first summation into $\alpha$ by setting $r=r_1N$, which can be viewed as splitting one $r$-rank LoRA into N sets (\eg, $r=4,N=2$ in Figure~\ref{fig:lora discussion}). In this way, we can get: 
\begin{equation}
\label{eq: ilora 3}
\begin{aligned}
    \Delta\Theta_2 = \sum_{j=1}^{r}\alpha_{j//N}B_jA_j,
\end{aligned}
\end{equation}
and $B_j,A_j$ are atom components of the $r$-rank LoRA. Therefore, in a decomposed view, we can see that the interaction between atom components remain partial and not sufficient (\ie, only $B_j$ and $A_j$ with the same $j$ have interactions), while the aggregation weight is customized by conventional recommendation model.

Built on these observations, we aim to enable more sufficient interaction between the atom components and extend the expressive power of LoRA parameters, with the integration of collaborative knowledge from CRM. To this end, we propose the component fully-interactive LoRA (CFLoRA) module. Intuitively, full interactions between atom components may have the following form:
\begin{equation}
\label{eq: CFLoRA decomposed}
\begin{aligned}
    \Delta\Theta_{CFLoRA} = \sum_{i=1}^{r}\sum_{j=1}^{r}w_{ij}B_iA_j,
\end{aligned}
\end{equation}
where each pair of $B_i,A_j$ can interact with each other, and $w_{ij}$ is generated by the CRM to control the interaction strength. In the next part, we will return to investigate LoRA in the matrices form, which is more concise and the basis of implementation.

\subsubsection{Composite View}
In the composite view, we return to analyze the LoRA matrices, instead of atom components. Revisiting Equation~\ref{eq: CFLoRA decomposed}, we can observe that it is mathematically consistent with the multiplication of several matrices:
\begin{equation}
\label{eq: CFLoRA, composite}
\begin{aligned}
    \Delta\Theta_{CFLoRA} = \sum_{i=1}^{r}\sum_{j=1}^{r}w_{ij}B_iA_j=BWA,
\end{aligned}
\end{equation}
where $W \in \mathbb{R}^{r\times r}$, and $w_{ij}$ fills the corresponding value in $W$. Correspondingly, we can rewritten Equation~\ref{eq: lora 1} and Equation~\ref{eq: ilora 3} in a similar form:
\begin{equation}
\label{eq: other methods, composite}
\begin{aligned}
    \Delta\Theta_1 &= \sum_{j=1}^r B_jA_j =BIA,\\
    \Delta\Theta_2 &= \sum_{j=1}^{r}\alpha_{j//N}B_jA_j =BW_2A,
\end{aligned}
\end{equation}
where $I\in \mathbb{R}^{r\times r}$ is the identity matrix. $W_2 \in \mathbb{R}^{r\times r}$ is a diagonal matrix where the diagonal elements can be divided into different $N$ blocks of different $\alpha$. A concrete example is shown in Figure~\ref{fig:lora discussion}, where $r=4, N=2$. The detailed matrices $I$, $W_2$ and $W$ are as follows:
\begin{equation}
    \label{eq: concrete W}
    I = 
    \begin{bmatrix}
    1 & 0 & 0 & 0 \\
    0 & 1 & 0 & 0 \\
    0 & 0 & 1 & 0 \\
    0 & 0 & 0 & 1 \\
    \end{bmatrix}
    , \quad
    W_2 = 
    \begin{bmatrix}
    \alpha_1 & 0 & 0 & 0 \\
    0 & \alpha_1 & 0 & 0 \\
    0 & 0 & \alpha_2 & 0 \\
    0 & 0 & 0 & \alpha_2 \\
    \end{bmatrix}
    , \quad
    W = 
    \begin{bmatrix}
    w_{11} & w_{12} & w_{13} & w_{14} \\
    w_{21} & w_{22} & w_{23} & w_{24} \\
    w_{31} & w_{32} & w_{33} & w_{34} \\
    w_{41} & w_{42} & w_{43} & w_{44} \\
    \end{bmatrix}.
\end{equation}


Comparing Equation~\ref{eq: CFLoRA, composite} and Equation~\ref{eq: other methods, composite}, we can observe that the way existing LLM4Rec methods employ LoRA are special cases of CFLoRA  in the composite view. CFLoRA can degrade into them with different constraints on $W$ (\eg, identity matrix or block diagonal matrix as Equation~\ref{eq: concrete W}). By analyzing LoRA parameters in both a composite view and a decomposed view, we provide a novel perspective to evaluate and compare these methods.

\subsubsection{CFLoRA Implementation} We have discussed that CFLoRA enables full interaction between atom components under the decomposed view, while preserving a concise matrix form under the composite view. Based on the matrix form, we can easily and elegantly implement the CFLoRA module. Specifically, for each data sample $x$, using the pretrained CRM, we can get the final representation $h$ according to Equation~\ref{eq: getting CRM representation}. With a lightweight projector (\eg, a two-layer MLP) and reshaping transformation, we can get the final $W$. The process can be formulated as:
\begin{equation}
\begin{aligned}
\label{eq: CFLoRA details}
h &= f(r_u, E_u, e_{i_c}) \in \mathbb{R}^{d_h}, \\
h_1 &= \operatorname{Projector}(h) \in \mathbb{R}^{r^2},\\
W &= \operatorname{Reshape}(h_1) \in \mathbb{R}^{r\times r}.
\end{aligned}
\end{equation}
In this way, CFLoRA extends the expressive capability of LoRA by enabling more comprehensive interaction, which makes the parameters effectively capture more useful information from the sequence. It can also be implemented easily and elegantly, with the integration of collaborative knowledge from CRM.

\subsection{Training and Inference Procedure}
We optimize the overall ReLLaX framework in an end-to-end manner, where the training objective is an extended version of causal language modeling in Equation~\ref{eq: causal LM} with the integration of ID modality inputs:
\begin{equation}
\max_{\Delta\Theta }\sum\nolimits_{(x^{ID},x^{text},y^{text})\in\mathcal{D}}\, \sum\nolimits_{j=1}^{|y^{text}|}\log P_{\Theta+\Delta\Theta}(y_{j}^{text}|x^{ID},x^{text},y_{<j}^{text}).
\end{equation}
The trainable parameters $\Delta\Theta$ now include the prompt generation projector in SPA, the component interaction projector in CFLoRA, as well as the vanilla LoRA $A,B$ matrices. The main weights of LLMs and the pretrained conventional recommendation model are frozen. 

Corresponding, the inference process of ReLLaX also needs to incorporate the ID modality inputs, changing Equation~\ref{eq: lora infer} into:
\begin{equation}
\begin{aligned}
    s &= \mathcal{M}_{\Theta+\Delta\Theta}(x^{ID},x^{text}).
\end{aligned}
\end{equation}

%% file: text/experiment.tex
\section{Experiment}

In this section, we conduct extensive experiments to answer the following research questions:
\begin{itemize}
    \item[\textbf{RQ1}] How does ReLLaX perform compared to existing baselines?
    \item[\textbf{RQ2}] What are the influences of different modules for ReLLaX?
    \item[\textbf{RQ3}] Does ReLLaX promote the lifelong sequential behavior comprehension ability of LLMs for recommendation tasks?

    \item[\textbf{RQ4}] How does ReLLaX help LLMs to better comprehend the user behavior sequence?
\end{itemize}


\subsection{Experiment Setup}
\subsubsection{Datasets}
We conduct experiments on three real-world datasets (\ie, BookCrossing\footnote{\url{http://www2.informatik.uni-freiburg.de/~cziegler/BX/}}, MovieLens-1M\footnote{\url{https://grouplens.org/datasets/movielens/1m/}} and MovieLens-25M\footnote{\url{https://grouplens.org/datasets/movielens/25m/}}) and show the dataset statistics in Table~\ref{tab:datasets}.
MovieLens-1M and MovieLens-25M datasets are split into training and testing sets with ratio of 8:1 according to the global timestamp~\cite{qin2021retrieval}. 
Since BookCrossing dataset has no timestamps, we follow previous work~\cite{bao2023tallrec} and  divide it into training and testing sets with ratio of 9:1 by random split of users.  Data samples with user behavior sequence length less than 5 are filtered on all three datasets. We describe more preprocessing details as follows:

\begin{itemize}
    \item \textbf{BookCrossing} possesses user-book integer ratings ranging from 0 to 10. We consider samples with rating above 5 as positive, and the rest as negative. 
    \item \textbf{MovieLens-1M} contains user-movie integer ratings ranging from 0 to 5. Samples with ratings of 4 and 5 are labeled as positive and the rest as negative.~\cite{zhou2018deep,xi2023towards}
    \item \textbf{MovieLens-25M} has a scoring range from 0 to 5, with increments of 0.5. We label samples with ratings above 3.0 as positive, and the rest as negative.
\end{itemize}



\begin{table}
    \caption{The dataset statistics.}
    \centering
    \resizebox{0.6\textwidth}{!}{
    \renewcommand\arraystretch{1.1}
    \begin{tabular}{c|cccccc}
    \toprule
     Dataset   & \#Users & \#Items & \#Samples & \#Fields & \#Features \\ 
     \midrule
     BookCrossing  & 278,858 & 271,375 & 17,714 & 10 & 912,279 \\
     MovieLens-1M & 6,040 & 3,706 & 970,009 & 10 & 16,944 \\
     MovieLens-25M & 162,541 & 59,047 & 25,000,095 & 6 & 280,576 \\ \bottomrule
    \end{tabular}
    }
    \label{tab:datasets}
\end{table}




\subsubsection{Evaluation Metrics}
To evaluate the recommendation performance of different models, we utilize AUC (area under the ROC curve), Log Loss (binary cross-entropy loss) and ACC (accuracy score) as the evaluation metrics. 
In the CTR prediction task, slightly higher AUC or lower Log Loss (e.g., 0.001) can be regarded as significant improvement~\cite{xDeepFM, DCNv2}.

\subsubsection{Baseline Models}
The CTR baseline models can be mainly classified into two categories: (1) \emph{conventional ID-based CTR models} that take one-hot encoded IDs as inputs, and (2) \emph{LM-based models} that incorporate pretrained language models and formulate CTR prediction as either text sequence-to-sequence or binary classification  problem.

Conventional ID-based CTR models can be further classified into two types: (1) feature interaction models, and (2) user behavior models. For feature interaction models, we choose DeepFM~\cite{DeepFM}, AutoInt~\cite{AutoInt}, and DCNv2~\cite{DCNv2} as representatives. For user behavior models, we choose  GRU4Rec~\cite{GRU4Rec}, Caser~\cite{Caser}, SASRec~\cite{SASRec}, DIN~\cite{zhou2018deep}, and SIM~\cite{SIM} as representatives.  
We perform average pooling over users' historical behaviors, and regard the outputs as additional feature fields for the feature interaction models.
SIM~\cite{SIM} is a classical CTR model that utilizes user behavior retrieval techniques to improve the recommendation performance. We include it for fair comparison, since ReLLaX incorporates semantic user behavior retrieval (SUBR).

As for LM-based CTR models, we select CTR-BERT~\cite{CTRBERT}, PTab~\cite{liu2022ptab}, P5~\cite{P5}, TALLRec~\cite{bao2023tallrec}, CoLLM~\cite{zhang2023collm}, iLoRA~\cite{kong2024customizing}, and the vanilla ReLLa~\cite{lin2024rella} as the representative baselines. CTR-BERT, PTab and P5 utilizes LMs of a small size, while TALLRec, CoLLM, iLoRA and ReLLa are based on LLMs and leverages LoRA-based instruction tuning.



\subsubsection{Implementation Details}

We select Vicuna-13B~\cite{vicuna2023} released by FastChat\footnote{\url{https://github.com/lm-sys/FastChat}} as the base LLM for ReLLaX and other LLM-based baselines.
All the experiments are conducted on V100 GPUs.
We adopt 8-bit quantization on the LLM weights for resource efficiency. The projectors leveraged in the SPA and CFLoRA modules are all two-layer MLPs with ReLU activation. The CFLoRA modules are applied on the query and value projection matrices of attention blocks. We follow previous works~\cite{bao2023tallrec,leng2023chinese-vicuna, lin2024rella} to set the configuration of the original LoRA decomposition matrices, with rank as 8, alpha as 16, and dropout as 0.05. The model is trained with a batch size selected from $\{128, 256\}$. The learning rate is initialized from $\{1\times 10^{-3}, 7 \times 10^{-4}\}$ with linear scheduler. On BookCrossing dataset, the maximum training epoch is set to 10, while on MovieLens-1M and MovieLens-25M datasets, the maximum epoch is set to 5. The configuration of baselines is in Appendix~\ref{app:baseline}.



Moreover, when constructing the prompt template for ReLLaX, we remove all the pure ID fields, \eg,  \textit{User ID}, \textit{Movie ID}, and \textit{Zipcode} fields on MovieLens-1M dataset.
The reason is that LLMs possess limited perceptual capabilities for pure ID texts~\cite{lin2023can}. Instead, we adapt the ID embeddings of corresponding items from the pretrained traditional CTR model as soft prompt, as described in Section~\ref{sec: SPA}. The detailed prompts are shown in Appendix~\ref{app:prompt}.
Note that we do not discard any features for other CTR baseline models, \ie, they take all the feature fields and user behavior sequences as inputs.


\subsection{Overall Performance (RQ1)}
\begin{table*}
\caption{The performance of different \textit{ID-based} and \textit{LM-based} models. 
The best result is given in bold, and the second-best value is underlined. 
\emph{Rel.Impr} denotes the relative AUC improvement rate of ReLLaX against each baseline. 
The symbol $\ast$ indicates statistically significant improvement of ReLLaX over the best baseline with $p$-value < 0.001.
}
\label{tab:overall performance}
\resizebox{\textwidth}{!}{
\renewcommand\arraystretch{1.1}
\begin{tabular}{c|c|cccc|cccc|cccc}
\toprule
\hline

\multicolumn{2}{c|}{\multirow{2}{*}{Model}} & \multicolumn{4}{c|}{BookCrossing} & \multicolumn{4}{c|}{MovieLens-1M} & \multicolumn{4}{c}{MovieLens-25M} \\ 
\multicolumn{2}{c|}{} & AUC  & Log Loss & ACC & Rel.Impr & AUC  & Log Loss & ACC & Rel.Impr & AUC  & Log Loss & ACC & Rel.Impr\\ 
   \hline 
   

\multicolumn{1}{c|}{\multirow{8}{*}{ID-based}} & DeepFM & 0.7496 & 0.5953 & 0.6760 & 1.05\% & 0.7915 & 0.5484 & 0.7225 & 1.49\% & 0.8189 & 0.4867 & 0.7709 & 3.52\% \\ 
\multicolumn{1}{c|}{\multirow{8}{*}{}} & AutoInt & 0.7481 & 0.6840 & 0.6365 & 1.26\% & 0.7929 & 0.5453 & 0.7226 & 1.31\% & 0.8169 & 0.4957 & 0.7689 & 3.77\% \\ 
\multicolumn{1}{c|}{\multirow{8}{*}{}} & DCNv2 & 0.7472 & 0.6816 & 0.6472 & 1.38\% & 0.7931 & 0.5464 & 0.7216 & 1.29\% & 0.8190 & 0.4989 & 0.7702 & 3.50\%\\ 
\multicolumn{1}{c|}{\multirow{8}{*}{}} & GRU4Rec & 0.7479 & 0.5930 & 0.6777 & 1.28\% & 0.7926 & 0.5453 & 0.7225 & 1.35\% & 0.8186 & 0.4941 & 0.7700 & 3.55\% \\ 
\multicolumn{1}{c|}{\multirow{8}{*}{}} & Caser & 0.7478 & 0.5990 & 0.6760 & 1.30\% & 0.7918 & 0.5464 & 0.7206 & 1.45\% & 0.8199 & 0.4865 & 0.7707 & 3.39\% \\ 
\multicolumn{1}{c|}{\multirow{8}{*}{}} & SASRec & 0.7482 & 0.5934 & 0.6811 & 1.24\% & 0.7934 & 0.5460 & 0.7233 & 1.25\% & 0.8187 & 0.4956 & 0.7691 & 3.54\% \\ 
\multicolumn{1}{c|}{\multirow{8}{*}{}} & DIN & 0.7477 & 0.6811 & 0.6557 & 1.31\% & 0.7962 & 0.5425 & 0.7252 & 0.89\% & 0.8190 & 0.4906 & 0.7716 & 3.50\% \\ 
\multicolumn{1}{c|}{\multirow{8}{*}{}} & SIM & 0.7541 & \textbf{0.5893} & 0.6777 & 0.45\% & 0.7992 & 0.5387 & 0.7268 & 0.51\% & 0.8344 & 0.4724 & 0.7822 & 1.59\% \\ 

\hline
\multicolumn{1}{c|}{\multirow{8}{*}{LM-based}}
& CTR-BERT & 0.7448 & 0.5938 & 0.6704 & 1.71\% & 0.7931 & 0.5457 & 0.7233 & 1.29\% & 0.8079 & 0.5044 & 0.7511 & 4.93\% \\ 
\multicolumn{1}{c|}{\multirow{8}{*}{}} & PTab & 0.7429 & 0.6154 & 0.6574 & 1.97\% & 0.7955 & 0.5428 & 0.7240 & 0.98\% & 0.8107 & 0.5022 & 0.7551 & 4.56\% \\ 
\multicolumn{1}{c|}{\multirow{8}{*}{}} & P5 & 0.7438 & 0.6128 & 0.6563 & 1.84\% & 0.7937 & 0.5478 & 0.7190 & 1.21\% & 0.8092 & 0.5030 & 0.7527 & 4.76\% \\ 
\multicolumn{1}{c|}{\multirow{8}{*}{}} & TALLRec & 0.7472 & 0.6144 & 0.6851 & 1.82\% & 0.7899 & 0.5473 & 0.7212 & 2.43\% & 0.8395 & 0.4540 & 0.7882 & 1.77\% \\
\multicolumn{1}{c|}{\multirow{8}{*}{}} & CoLLM & 0.7482 & 0.5965 & 0.6834 & 1.68\% & 0.7948 & 0.5454 & 0.7232 & 1.80\% & 0.8471 & \underline{0.4453} & 0.7921 & 0.86\% \\
\multicolumn{1}{c|}{\multirow{8}{*}{}} & iLoRA & 0.7514 & 0.6162 & \textbf{0.6902} & 1.25\% & 0.8022 & \underline{0.5354} & \underline{0.7283} & 0.86\% & 0.8435 & 0.4499 & 0.7907 & 1.29\% \\
\multicolumn{1}{c|}{\multirow{5}{*}{}} & ReLLa & 0.7575 & 0.5919 & 0.6806 & 0.44\% & \underline{0.8033} & 0.5362 & 0.7280 & 0.72\% & \underline{0.8477} & 0.4524 & \underline{0.7925} & 0.79\% \\ 
\multicolumn{1}{c|}{\multirow{8}{*}{}} & ReLLaX & \textbf{0.7608$^*$} & \underline{0.5896} & \underline{0.6845} & - & \textbf{0.8091$^*$} & \textbf{0.5301$^*$} & \textbf{0.7355$^*$} & - & \textbf{0.8544$^*$} & \textbf{0.4369$^*$} & \textbf{0.7988$^*$} & - \\ 
   \hline  
   \bottomrule          
\end{tabular}
}
\end{table*}

We evaluate the performance of ReLLaX in comparison to existing baseline models, and report the results in Table~\ref{tab:overall performance}.
Note that as the training of LLM-based models (\ie, TALLRec, CoLLM, iLoRA, ReLLa and ReLLaX) is both time-consuming and resource-intensive,  we follow previous works~\cite{lin2024rella, zhu2024lifelong, bao2023tallrec} and train LLM-based baseline models with a \textit{few-shot} strategy, where 2048/65536/65536 training samples are randomly sampled from the full training set on BookCrossing/MovieLens-1M/MovieLens-25M, respectively. For other baseline models, we train them with the entire training set. The length of user behavior sequence for all models is fixed to 60/30/30 for the three datasets respectively.

From Table~\ref{tab:overall performance} we can have the following observations:

\begin{itemize}
    \item ReLLaX achieves significant performance improvement above the traditional ID-based CTR models. It is worth noting that theses ID-based CTR models are trained on the entire training set, while ReLLaX only utilizes less than 10\% training samples for finetuning. This demonstrates the superior data efficiency of ReLLaX for sequential recommendation tasks. 

    \item 
    ReLLaX performs better than LLM-based baseline methods utilizing different data-level and prompt-level enhancement strategies. Among these methods, CoLLM uses traditional CTR model to generate soft prompt but overlooks the general data quality. In comparision, ReLLa utilizes retrieval to improve the data quality but fails to utilizes collaborative knowledge explicitly. Unlike these methods, ReLLaX adopts a full-stack optimization framework, providing both data-level and prompt-level enhancement. ReLLaX not only improves the data quality of pure textual prompt via SUBR, but also but also make the item representations in the language space more aligned with recommendation through SPA, further reducing the difficulty of reasoning over the user behavior sequences.

    \item  
    ReLLaX generally outperforms LLM-based baseline methods which employs LoRA in different ways, validating the efficacy of our proposed component fully-interactive LoRA (CFLoRA) module for parameter-level enhancement. As illustrated in Section~\ref{sec:parameter-level enhancement}, in the composite view, the ways these methods employ LoRA are essentially degraded versions of our proposed CFLoRA with different constraints, while CFLoRA imposes no constraints on the matrices. In the decomposed view, our CFLoRA module is more general and enables more sufficient interaction between LoRA atom components, extending the expressive capability of LoRA and leading to better recommendation performance.



\end{itemize}

\begin{table*}
\caption{ The performance of different variants of ReLLaX. We remove different modules of ReLLaX to evaluate the contribution of each part to the model. 
The best result is given in bold, and the second-best value is underlined.
}
\label{tab:ablation}
\resizebox{0.9\textwidth}{!}{
\renewcommand\arraystretch{1.2}
\begin{tabular}{c|ccc|ccc|ccc}
\toprule
\hline

\multicolumn{1}{c|}{\multirow{2}{*}{Model Variant}} & \multicolumn{3}{c|}{BookCrossing} & \multicolumn{3}{c|}{MovieLens-1M} & \multicolumn{3}{c}{MovieLens-25M} \\ 
\multicolumn{1}{c|}{} & AUC  & Log Loss & ACC & AUC  & Log Loss & ACC & AUC  & Log Loss & ACC \\ 
   \hline 

ReLLaX (Ours) & \textbf{0.7525} & \underline{0.6077} & \underline{0.6845} & \textbf{0.7978} & \textbf{0.5423} & \textbf{0.7281} & \textbf{0.8343} & \textbf{0.4724} & \textbf{0.7859} \\
ReLLaX (w/o SUBR) & \underline{0.7520} & 0.6180 & \textbf{0.6867} & \underline{0.7963} & \underline{0.5467} & \underline{0.7252} & \underline{0.8335} & \underline{0.4759} & \underline{0.7836} \\
ReLLaX (w/o SPA) & 0.7507 & 0.6098 & 0.6828 & 0.7932 & 0.5586 & 0.7169 & 0.8298 & 0.4766 & 0.7815 \\
ReLLaX (w/o CFLoRA) & 0.7498 & 0.6339 & 0.6733 & 0.7925 & 0.5515 & 0.7177 & 0.8288 & 0.4771 & 0.7784 \\
ReLLa & 0.7399 & \textbf{0.6002} & 0.6715 & 0.7849 & 0.5693 & 0.6985 & 0.8192 & 0.4904 & 0.7715 \\ 
ReLLa (w/o IT) & 0.7253 & 0.9277 & 0.5750 & 0.7013 & 0.6250 & 0.6507 & 0.7324 & 0.5858 & 0.7027 \\ 
ReLLa (w/o IT \& SUBR) & 0.7176 & 0.9507 & 0.5649 & 0.6993 & 0.6291 & 0.6493 & 0.7503 & 0.6308 & 0.6427 \\ 
  
   \hline  
   \bottomrule          
\end{tabular}
}
\end{table*}

\subsection{Ablation Study (RQ2)}
\label{sec:ablation}

To analyze the efficacy of each module in our proposed ReLLaX framework, we design several model variants of ReLLaX. The number of training samples are fixed to 1024/8192/8192, and the length of textual user behavior sequnce is set to $60/30/30$ for BookCrossing/MovieLens-1M/MovieLens-25M datasets, respectively. The length of user behavior sequences for the pretrained traditional CTR model is set to 60. The designed model variants of ReLLaX are as follows:
\begin{itemize}
    \item \textbf{ReLLaX (Ours)} is the complete version of our proposed method, with SUBR, SPA and CFLoRA providing data-level, prompt-level and parameter-level enhancement respectively.
    \item \textbf{ReLLaX (w/o SUBR)}. We remove the semantic user behavior retrieval for both training and testing samples. That is, training and testing data uses prompts without retrieval enhancement. This variant aims to ablate on the efficacy of data augmentation brought by retrieval.
    \item \textbf{ReLLaX (w/o SPA)}. We remove the soft prompt augmentation for both training and testing samples. That is, the prompts are the same as vanilla ReLLa, without explicitly injecting collaborative knowledge into the prompt.
    \item \textbf{ReLLaX (w/o CFLoRA)}. We remove CFLoRA modules. Therefore, the instruction tuning utilizes vanilla LoRA, with SUBR to improve the data quality and SPA to generate soft prompt.
    \item \textbf{ReLLa}. This variant corresponds to removing all the new proposed techniques in this paper, \ie\ SPA and CFLoRA.
    \item \textbf{ReLLa (w/o IT)}. We remove the instruction tuning from ReLLa, while still leveraging SUBR. This variant indicates the zero-shot setting of ReLLa, which aims to ablate on the efficacy of data augmentation of SUBR under the zero-shot setting.
    \item \textbf{ReLLa (w/o IT \& Retrieval)}. We remove both the instruction tuning and retrieval operation. Therefore, the testing data only contains original data samples. This variant indicates the zero-shot version of vanilla Vicuna-13B.
\end{itemize}

The performance of these variants are presented in Table~\ref{tab:ablation}, from which we can draw the following observations:
\begin{itemize}
    \item Removing each module from ReLLaX generally leads to performance degradation, while these variants of ReLLaX still outperform the vanilla ReLLa, demonstrating the efficacy of each component proposed in this paper. Moreover, ReLLaX (w/o SUBR) is generally second only to ReLLaX (Ours), which indicates that the proposed SPA and CFLoRA contribute more to the performance gain than SUBR. This highlights the importance of the injection of collaborative knowledge from traditional CTR model. Specifically, SPA generates soft prompt from CTR model to reduce the difficulty of user sequence modeling, while CFLoRA extends LoRA parameters' expressive ability to capture long-sequence information more effectively.
    \item 
    Comparing ReLLa (w/o IT) and ReLLa (w/o IT \& Retrieval), which fall back into zero-shot settings, we can observe that ReLLa (w/o IT) generally achieves significant improvements over ReLLa (w/o IT \& Retrieval), except for the AUC metric on MovieLens-25M.
    This demonstrates that semantic user behavior retrieval (SUBR) serves as data augmentation and improves the quality of data samples, making the filtered behavior sequence more friendly for LLM to extract useful knowledge.
\end{itemize}

\begin{figure*}[t]
\centering
\includegraphics[width=0.97\textwidth]{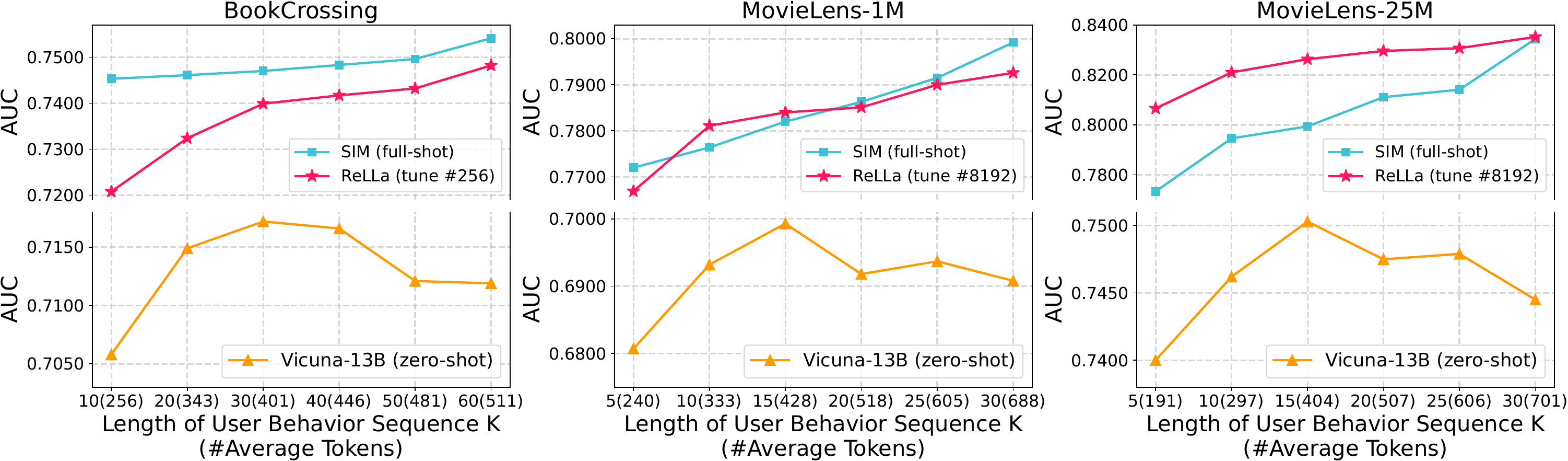}
  \caption{The AUC performance of different models w.r.t. different length of user behavior sequence $K$. 
  ReLLa (\ie, ReLLaX w/o SPA and CFLoRA) manages to alleviate the incomprehension problem of LLMs on recommendation tasks with long textual user behavior sequences.
  }
  \label{fig:vary K}
\end{figure*}

\subsection{Sequential Behavior Comprehension (RQ3)}
\label{sec:study on shot K}


We vary the length of user behavior sequence to analyze its impact on CTR prediction performance, which can reflect the comprehension ability of a model towards user behavior sequences. It is noteworthy that ReLLaX utilizes user behavior sequences of two modalities, \eg\ text modality in the hard prompt, and ID modality for CRM in the SPA and CFLoRA module. As the length of user behavior sequences increases, the number of tokens in the textual sequence grows, leading to more VRAM consumption. In comparison, as shown in Equation~\ref{eq: CFLoRA details}, sequences in ID modality can be encoded into dense representations, and will incur no additional resource overhead as the sequence length increases. Therefore, in this section, we first vary the length of textual behavior sequence $K$ with vanilla ReLLa, which removes the ID behavior sequences from ReLLaX. Finding the best and largest $K$ that computation resource allows, we bring back the ID modality, fix the largest $K$ and evaluate ReLLaX with ID behavior sequences with different length $L$.

\subsubsection{Different textual sequence length for ReLLa}
\label{sec: ReLLa K}

We evaluate ReLLa (few-shot), which uses the pure textual behavior sequences, against SIM (full-shot) and Vicuna-13B (zero-shot) using the same sequence length. The number of training examples (\ie, shots) for ReLLa is 256/8192/8192 for BookCrossing, MovieLens-1M and MovieLens-25M respectively, while SIM (full-shot) is trained with the entire training set and Vicuna-13B (zero-shot) is not tuned. In our experiment setting, the maximum tokens allowed by the resource is about 700 tokens. Hence, we choose the largest textual sequence length $K$ to 60/30/30 for the three datasets respectively, and evaluate the performance within the largest $K$. The results are illustrated in Figure~\ref{fig:vary K}, from which we can have the following observations:
\begin{itemize}
    \item As a conventional recommendation model, SIM (full-shot) enjoys steady performance improvement when the ID sequence length $K$ increases. This is consistent with our common knowledge, where longer user behavior sequences can provide more useful information to better accomplish the CTR prediction task. 
    \item Nevertheless, the performance of Vicuna-13B (zero-shot) only peaks at $K=30/15/15$ on BookCrossing/MovieLens-1M/MovieLens-25M datasets respectively, and then starts to decrease with longer sequence. It is noteworthy that the number of involved tokens with the largest $K$ (\ie, about 700 tokens) is actually far from reaching the context window limitation of Vicuna-13B (\ie, 2048 tokens). 
    This indicates that it is non-trivial for LLMs to comprehend the pure  textual context of long behavior sequences in recommendation, which requires a certain amount of in-domain knowledge.
    \item ReLLa alleviates the incomprehension problem of LLMs on long user behavior sequences for recommendation. 
    Compared with Vicuna-13B (zero-shot), whose performance drops when $K>30$ on BookCrossing and $K>15$ on MovieLens-1M and MovieLens-25M, there are no performance turning points for ReLLa. 
    Similar to SIM (full-shot), the AUC performance of ReLLa (few-shot) achieves continuous improvement as $K$ increases, validating the enhanced comprehension ability of ReLLa for the pure textual contexts with longer behavior sequences.
\end{itemize}

\subsubsection{Further longer ID sequence length for ReLLaX}

\begin{figure}[t]
\centering
\includegraphics[width=0.95\textwidth]{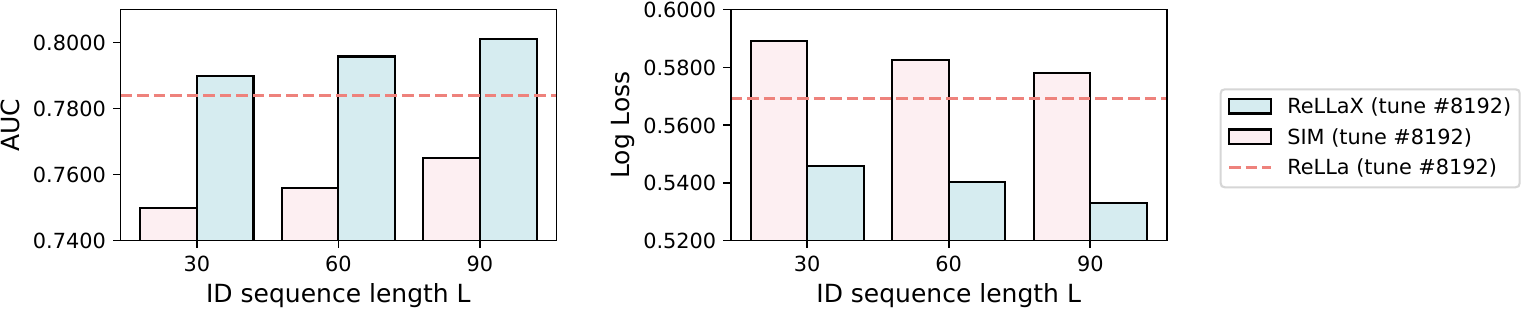}
  \caption{
The performance of ReLLaX w.r.t. different $L$ on the MovieLens-1M dataset. The $L$ is exactly the ID sequence length to pretrain the conventional recommendation model used in ReLLaX. The textual sequence length for the prompt of ReLLa and ReLLaX is fixed to 30, which is the best and largest value allowed by the computation resource. 
  }
  \label{fig:ctr K}
\end{figure}


In section~\ref{sec: ReLLa K}, we have found that under the computation resource limitation and a mediate $K$ (\ie, $K$<=30), ReLLa can enjoy better CTR prediction performance as $K$ increases, thus mitigating the  behavior sequence incomprehension problem. Here, we fix textual sequence length $K$ to the maximum (\ie, 30 on the MovieLens-1M dataset). By further introducing SPA and CFLoRA to ReLLa, we evaluate ReLLaX with different ID sequence length $L$, to see if further longer ID sequence can improve the understanding of the long textual sequence (\ie, $L>=K$). The number of training samples for the models is 8192. The results are reported in Figure~\ref{fig:ctr K}, and we can observe that:
\begin{itemize}
    \item Injecting collaborative knowledge into the LLM by SPA and CFLoRA can improve the recommendation performance. Even when the ID sequence length $K$ is equal to the text sequence length (\ie, $L=30, K=30$), ReLLaX can still outperform ReLLa, while the exact items composing the sequences are the same in the text and ID sequence. This indicates that the soft prompt injected using the SPA  module can improve the understanding of the pure textual behavior sequences, and CFLoRA is more effective to capture long-sequence information than vanilla LoRA in ReLLa.
    \item ReLLa and ReLLaX are more sample-efficient than SIM. With the same number of training samples, ReLLa and ReLLaX can achieve better recommendation performance.
    \item The performance of ReLLaX improves as $L$ increases. This can be attributed to that more useful information is encoded and utilized by the LLM as $L$ grows, thus further enhancing the comprehension ability over the behavior sequences.
\end{itemize}

\begin{figure}[t]
\centering
\includegraphics[width=0.75\textwidth]{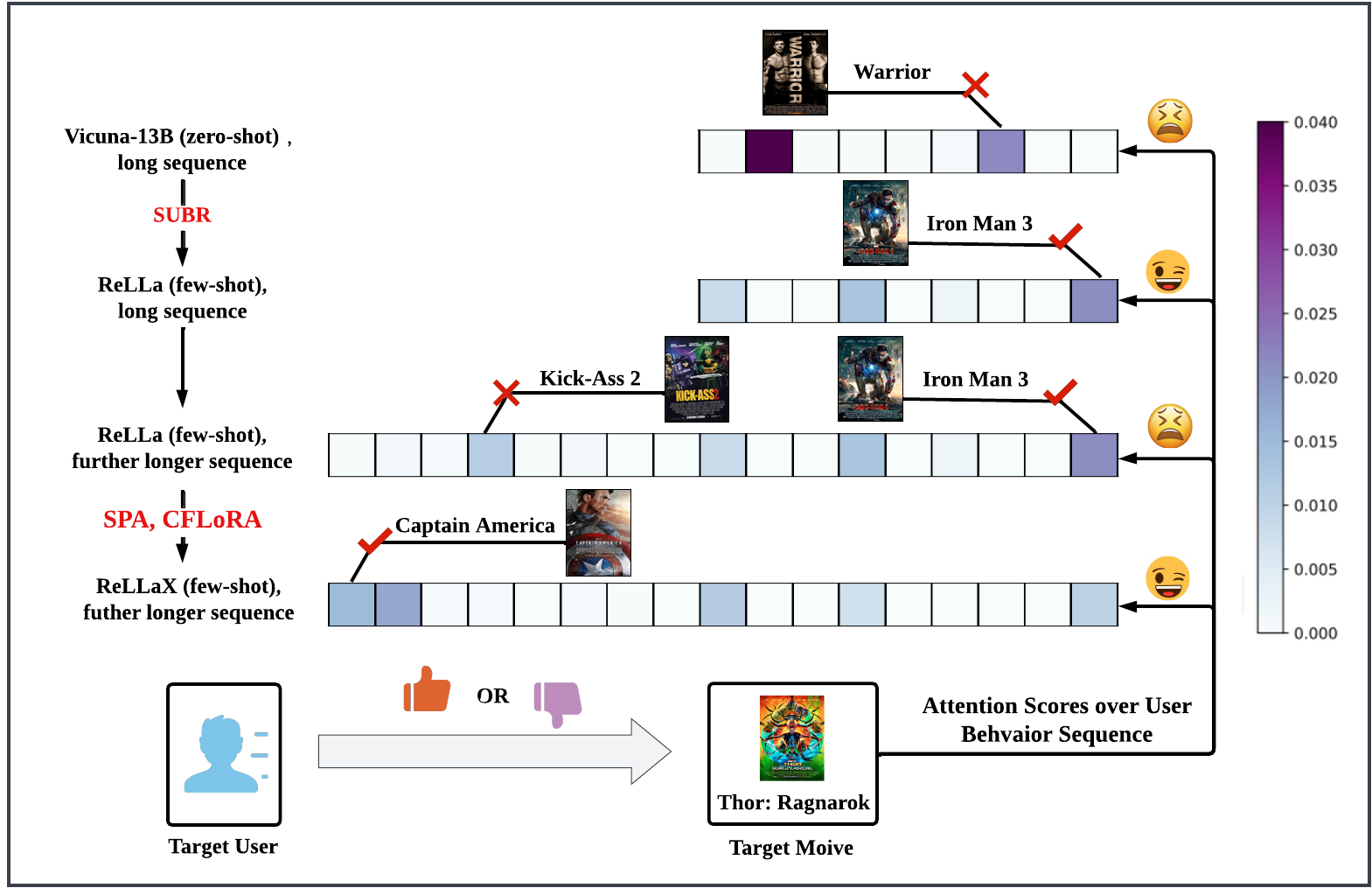}
  \caption{
Case study of ReLLaX on the MovieLens-25M dataset.
We visualize the attention scores for the items that users have interacted with (represented by rectangles) at the last hidden layer of the LLM. The deeper the  color of a rectangle, the higher the attention score of the corresponding historical item, indicating its greater contribution to the final CTR estimation.
  }
  \label{fig:case study}
\end{figure}


\subsection{Case Study (RQ4)}
\label{sec: case study}

In this section, we perform a case study to explore how ReLLaX enhances the ability of LLMs to understand long user behavior sequences. As depicted in Figure~\ref{fig:case study}, we choose a testing sample from the MovieLens-25M dataset, and visualize the attention scores of the target item across the user behavior sequence for three models, \ie\ Vicuna-13B (zero-shot), ReLLa (few-shot), and ReLLaX (few-shot). The attention score for each historical item is calculated by summing up the attention scores of each word token for the corresponding item at the last hidden layer of the LLM. In Figure~\ref{fig:case study}, each historical item is represented by a rectangle, with colors ranging from blue to purple. The deeper the color, the higher the attention score of the corresponding item, indicating a greater contribution to the final CTR estimation.

On the long sequence, for Vicuna-13B (zero-shot), the movie \emph{Warrior} is given high attention, which have little relevance to the target movie \textit{Thor: Ragnarok}. Consequently, the model fails to accurately predict the user's preference for the target item. By incorporating semantic user behavior retrieval (SUBR), we can homogenize the sequence and introduce more relevant items. As shown in Figure~\ref{fig:case study}, ReLLa (few-shot) focuses more on superhero movies like \emph{Iron Man 3}, which are semantically similar to the target item. 

However, although SUBR can lower the heterogeneity score, as the length of sequence increases further, the heterogeneity of the longer sequence can still raise to a high level where LLMs fail to comprehend it, which is also illustrated in Table~\ref{tab:heter score}. Hence, in Figure~\ref{fig:case study}, with further longer sequences, there are still some irrelevant items, such as \emph{Kick-Ass 2}, which is not closely related to \textit{Thor: Ragnarok} produced by Marvel. By further implementing soft prompt augmentation (SPA) and instruction tuning with component fully-interactive LoRA (CFLoRA), ReLLaX (few-shot) demonstrates a more refined attention mechanism. The significant attention weights are now concentrated on relevant superhero movies produced by Marvel. This indicates that the techniques we propose can help LLMs accurately identify the correlation between the target item and historical items, thereby improving their understanding of user behavior sequences.

%% file: text/related_work.tex
\section{Related Work}
\label{sec: related work}

\subsection{Traditional CTR Prediction}

Click-through rate (CTR) prediction is a critical component in a variety of online applications, such as recommender systems~\cite{xi2023bird}, advertising~\cite{ou2023deep}, and web search~\cite{lin2021graph, fu2023f, dai2021adversarial}. Its objective is to accurately estimate the likelihood of a user clicking on a particular target item within a given context~\cite{zhang2021deep,lin2023map,wang2024flip}. Traditional CTR prediction models can be broadly categorized into two groups: (1) feature interaction-based models and (2) sequential recommendation models.

Feature interaction-based models primarily draw inspiration from POLY2~\cite{POLY2} and FM~\cite{FM}, aiming to capture higher-order interactions between features across multiple fields using various operators (e.g., product~\cite{PNN,DCNv2,DeepFM,EDCN}, convolution~\cite{CFM,FGCNN}, and attention~\cite{AutoInt,AFM}). For example, DCN~\cite{DCNv1}, xDeepFM~\cite{xDeepFM}, and DCNv2~\cite{DCNv2} employ product-based feature crossing operations at each layer to explicitly model higher-order feature interactions. Meanwhile, AutoInt~\cite{AutoInt} and InterHAt~\cite{InterHAt} use attention mechanisms to model feature interactions, which also provide interpretable predictions via attention weights.

Sequential recommendation models~\cite{zhou2019deep, pi2019practice, zhou2018deep} focus on modeling user behavior and aim to dynamically capture a user's interests toward a target item based on their behavioral history. These models utilize different architectures, such as recurrent neural networks (RNNs)\cite{hidasi2017recurrent, GRU4Rec}, convolutional neural networks (CNNs)\cite{Caser}, attention~\cite{zhou2019deep, zhou2018deep}, and memory banks~\cite{pi2019practice, ren2019lifelong}, to model user behavior sequences and infer user preferences. For example, GRU4Rec~\cite{GRU4Rec} uses gated recurrent units (GRU)\cite{chung2014empirical} to encode sequential user behaviors, while Caser\cite{Caser} introduces CNNs to model higher-level patterns in user behavior sequences.

\subsection{Language Models for Recommendation}

Recent studies~\cite{lin2023can} have explored the use of language models in recommendation systems, categorizing their roles within the recommendation pipeline: (1) feature engineering~\cite{liu2023first, borisov2022language, li2023taggpt, mysore2023large, carranza2023privacy, christakopoulou2023large, shan2024automatic, wang2024learnable}, (2) feature encoding~\cite{muhamed2021ctr, hou2022towards, yu2021tiny, wang2022transrec, hou2023learning, zhang2022twhin, fu2023exploring, yuan2023go, qiu2021u, li2023exploring}, and (3) scoring/ranking functions~\cite{liu2022ptab, kang2023llms, zhang2021language, li2023pbnr, bao2023tallrec, li2023text, zhang2023prompt, mao2023unitrec, hua2023up5, geng2023vip5, hua2023index, zhang2023chatgpt, hou2023large, chen2023palr, petrov2023generative, wang2023zero}.

In feature engineering, large language models (LLMs) process raw data (e.g., user profiles and item descriptions) and generate additional textual features through carefully designed prompts and templates. For example, KAR~\cite{xi2023towards} uses LLMs to capture user preferences and item-related knowledge through factorization prompting techniques, providing augmented features that improve recommendation performance in a model-agnostic manner. Similarly, GENRE~\cite{liu2023first} employs LLMs to generate news summaries, synthetic news pieces, and user profiles.

For feature encoding, LLMs are utilized as auxiliary encoders to enrich user/item representations with semantic information and facilitate cross-domain recommendation through natural language interfaces. For instance, U-BERT~\cite{qiu2021u} enhances user representations by encoding review texts into dense vectors using BERT. Similarly, UniSRec~\cite{hou2022towards} and VQ-Rec~\cite{hou2023learning} leverage BERT to encode item descriptive texts, enabling unified cross-domain sequential recommendation.

In the context of scoring and ranking, researchers have explored the potential of LLMs to serve directly as core modules for item scoring or ranking, rather than as assistants to traditional recommendation models. These studies have used LLMs to perform tasks such as item scoring~\cite{liu2022ptab, kang2023llms, zhang2021language, li2023pbnr, bao2023tallrec, li2023text, zhang2023prompt, mao2023unitrec, bao2024real, ren2024self}, or item generation~\cite{hua2023up5, geng2023vip5, hua2023index, zhang2023chatgpt, hou2023large, chen2023palr, petrov2023generative, wang2023zero}. Furthermore, some works~\cite{geng2022recommendation, cui2022m6, zhang2023recommendation, liu2023chatgpt, sun2023chatgpt, dai2023uncovering} utilize the multi-task capabilities of LLMs to address multiple tasks (e.g., both scoring and generation) through a unified language interface. Recently, a new approach of using LLM to predict scores together with traditional CTR models rises~\cite{zhang2023collm,   zhu2024lifelong, kong2024customizing,liao2024llara}, balancing both the open-world knowledge of LLM and the collaborative filtering knowledge of traditional CTR models.

This paper primarily focuses on leveraging LLMs as the scoring/ranking function, using pointwise scoring for CTR prediction. To the best of our knowledge, we are the first to identify and systematically address the issue of LLMs' inability to comprehend lifelong user behavior sequences when applied to scoring and ranking tasks. We propose a novel ReLLaX framework to mitigate this challenge by incorporating retrieval techniques from the data perspective, utilizing soft prompt augmentation from prompt perspective, and let large language models fully cooperate with conventional recommendation models from parameter perspective, thereby enhancing the LLMs' ability to comprehend and improve their recommendation performance.

%% file: text/conclusion.tex
\section{Conclusion}

In this paper, we focus on adapting and enhancing LLMs as the scoring/ranking function for recommendation tasks.
We first identify and formulate the lifelong behavior sequence incomprehension problem of LLMs, \ie, LLMs fail to extract useful information from a textual context of long user behavior sequence, even if the length of context is far from reaching the context limitation of LLMs. To address such an issue and promote the recommendation performance , we propose a novel ReLLaX framework, which presents full-stack optimization for current LLM4Rec paradigm. Specifically, we design semantic user behavior retrieval (SUBR), soft prompt augmentation (SPA) and component fully-interactive LoRA (CFLoRA) to provide data-level, prompt-level and parameter-level enhancement, respectively. We also present a new perspective to compare and analyze existing LoRA-based LLM4Rec methods, where LoRA parameters are revisited under both a composite and a decomposed view. Extensive experiments validate the effectiveness of our proposed ReLLaX compared with existing baselines.

Although ReLLaX demonstrates outstanding performance and sample efficiency compared to traditional ID-based models in sequential recommendation, we have to acknowledge that its inference speed is slower than these ID-based models. Therefore, ReLLaX is currently better suited for real-world applications where latency tolerance is high, such as conversational recommendation or conversational search. However, this computational limitation is inherent to the large-scale nature of LLMs and is not specific to ReLLaX. Rather, it is a broader challenge within the research community working on LLMs for recommendation systems. Our future work may focus more on enhancing inference efficiency through techniques like pruning and distillation. Additionally, improving training efficiency is another area where LLM4Rec methods, including ReLLaX, can progress.

%% file: text/appendix.tex
\appendix

\section{Prompt Illustration}
\label{app:prompt}

We demonstrate several examples to illustrate the hard prompt templates used for ReLLaX on all three datasets.  

\begin{figure}[t]
\centering
\includegraphics[width=0.55\textwidth]{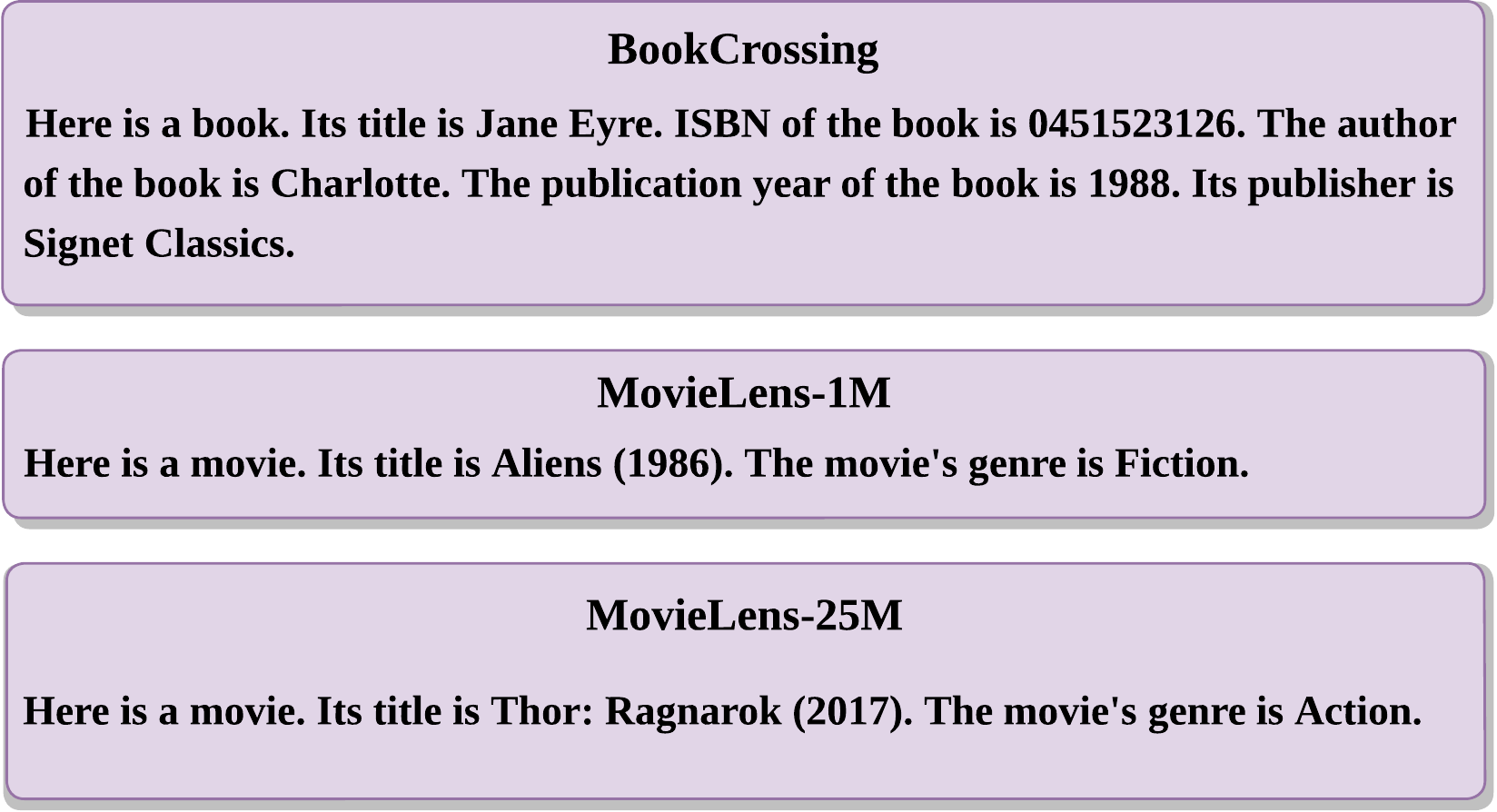}
  \caption{Examples of hard prompt templates of item descriptions for three datasets. The textual description is used to obtain the semantic item embedding from LLM, which will then be leveraged by SUBR.
  }
  \label{fig: full item description}
\end{figure}

Figure~\ref{fig: full item description} demonstrates how we design prompts for item descriptions on the three datasets, which will be encoded by LLM for semantic user behavior retrieval (SUBR). 

\begin{figure}[t]
\centering
\vspace{-7pt}
\includegraphics[width=0.55\textwidth]{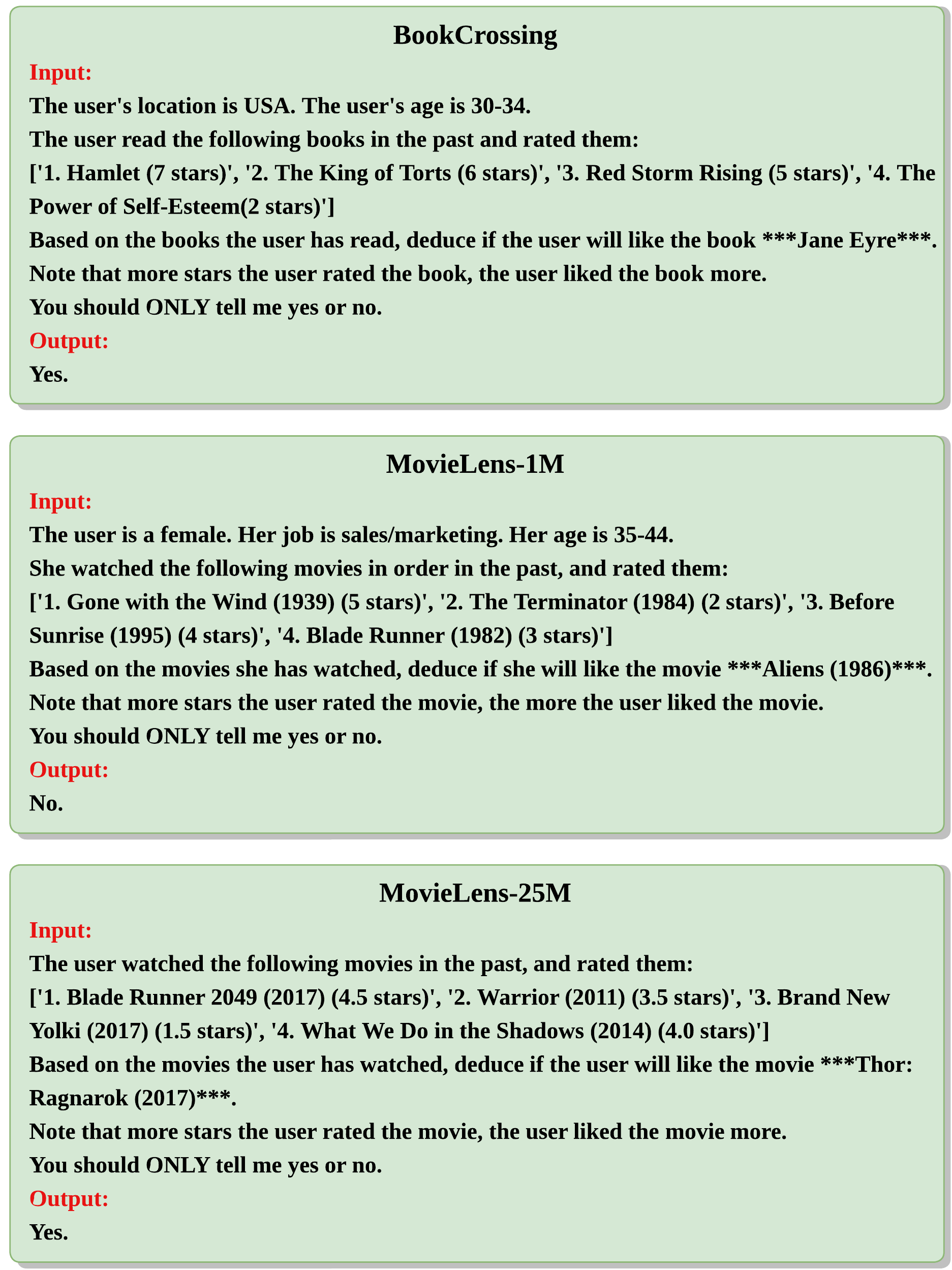}
  \caption{Examples of vanilla prompt templates for three datasets \emph{without} SUBR. The user behavior sequence is constructed by the most recent $K$ items.
  }
  \vspace{-10pt}
  \label{fig: prompt simple}
\end{figure}

\begin{figure}[t]
\centering
\vspace{-7pt}
\includegraphics[width=0.55\textwidth]{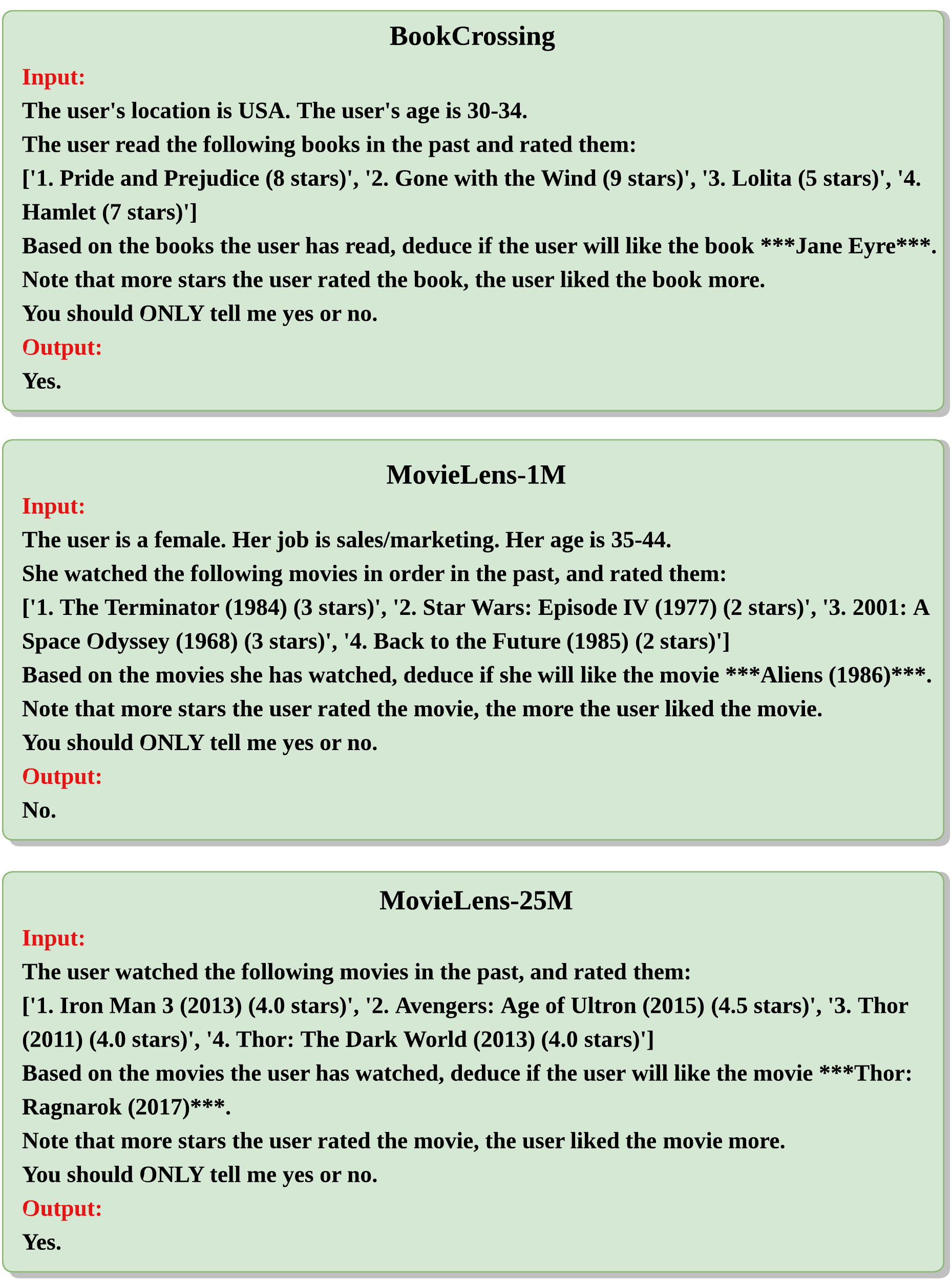}
  \caption{Examples of retrieval-enhanced prompt templates for three datasets \emph{with} SUBR. The user behavior sequence is constructed by the most relevant $K$ items.
  }
  \label{fig: prompt ret}
\end{figure}

\begin{figure}[t]
\centering
\vspace{-7pt}
\includegraphics[width=0.55\textwidth]{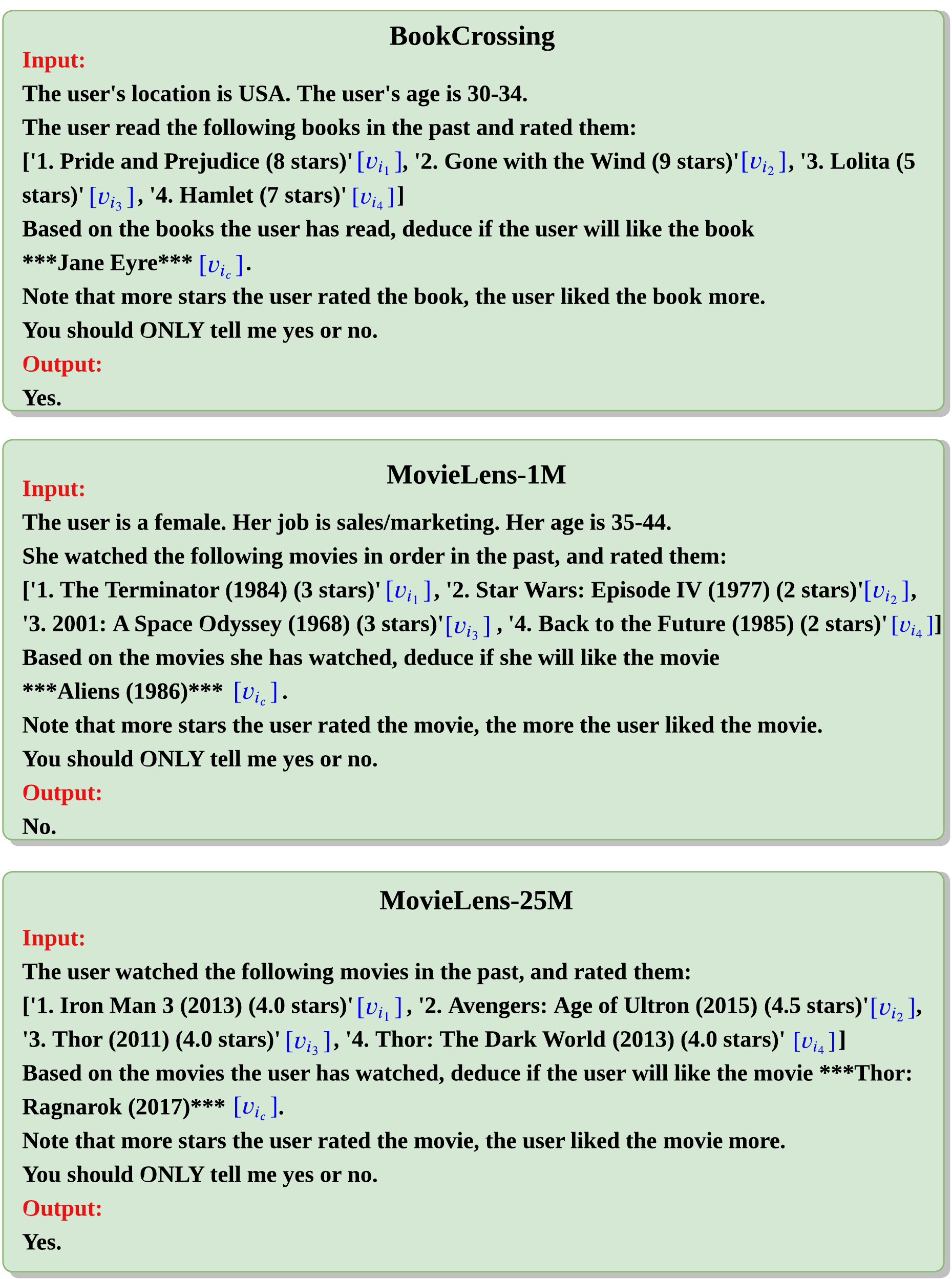}
  \caption{Examples of comprehensive prompt templates for three datasets \emph{with} SUBR and SPA. The user behavior sequence is constructed by the most relevant $K$ items, and the blue tokens are soft prompts generated from the item ID embeddings of the CRM .
  }
  \label{fig: prompt final}
\end{figure}

Figure~\ref{fig: prompt simple} shows the textual input-output pairs without semantic user behavior retrieval (SUBR), where the user behavior sequence is truncated to most recent $K$ (\eg, $K=4$ in the figure). 
As is shown in Figure~\ref{fig: prompt ret}, after applying SUBR, the user behavior sequence will be replaced by most relevant $K$ historical items towards the target item.
For example, for MovieLens-25M dataset, historical behaviors retrieved by SUBR are all related to superheros or Marvel, which is highly correlated to the target movie ``Thor: Ragnaro''. 
Note that the user behavior sequence generated by SUBR keeps the chronological  order in the original lifelong user sequence. Furthermore, employing soft prompt augmentation (SPA), Figure~\ref{fig: prompt final} shows the final prompts fed into the LLM, where we insert soft prompts (represented by the blue tokens) into the prompt, which are generated from item ID embeddings from CRMs.

\section{Baseline Implementation}
\label{app:baseline}

In this section, we describe the hyperparameter configuration for the baseline models from two different categories: (1) traditional CTR models, and (2) LM-based models.

\subsection{Traditional CTR Models}
We have conducted hyperparameter tuning for various models on different datasets. For the BookCrossing dataset, the embedding size is chosen from \{8, 16, 32\}, while for the MovieLens-1M and MovieLens-25M datasets, it is selected from \{32, 64\}. The dropout rate is set to one of \{0.0, 0.1, 0.2\}. ReLU is utilized as the activation function. A learning rate of $1\times 10^{-3}$ is applied, and the AdamW optimizer is employed. Regarding batch size, on the BookCrossing dataset, it is picked from \{32, 64\}, whereas on the MovieLens-1M and MovieLens-25M datasets, it is determined from \{256, 512\}. Additional model-specific hyperparameter settings are as follows:

\begin{itemize}[leftmargin=10pt]
    \item \textbf{DeepFM}~\cite{DeepFM}. For the BookCrossing dataset, the DNN layer size is determined from \{32, 64, 128\}, and the number of DNN layers is chosen from \{1, 2, 3\}. On the MovieLens-1M and MovieLens-25M datasets, the DNN layer size is selected from \{128, 256\}, and the number of DNN layers is set to one of \{3, 6, 9, 12\}.
    \item \textbf{AutoInt}~\cite{AutoInt}. On the BookCrossing dataset, the number of attention layers is selected from \{1, 2\}, with the attention size fixed at 32. For the MovieLens-1M and MovieLens-25M datasets, the number of attention layers is chosen from \{3, 6, 9, 12\}, and the attention size is selected from \{64, 128, 256\}. The number of attention heads is consistently set to 1.
    \item \textbf{DCNv2}~\cite{DCNv2}. On the BookCrossing dataset, the DNN layer size is picked from \{32, 64, 128\}, and both the number of DNN layers and cross layers are selected from \{1, 2, 3\}. For the MovieLens-1M and MovieLens-25M datasets, the DNN layer size is determined from \{128, 256\}, and the number of DNN layers and cross layers are chosen from \{3, 6, 9, 12\}.
    \item \textbf{GRU4Rec}~\cite{GRU4Rec}. The number of GRU layers is selected from \{1, 2, 3\}. On the BookCrossing dataset, both the GRU hidden size and DNN hidden size are chosen from \{32, 64\}. For the MovieLens-1M and MovieLens-25M datasets, the GRU hidden size and DNN hidden size are selected from \{64, 128, 256\}.
    \item \textbf{Caser}~\cite{Caser}. The number of vertical convolution kernels is selected from \{2, 4, 8\}, and the number of horizontal convolution kernels is determined from \{4, 8, 16\}. The number of DNN layers is picked from \{1, 2, 3\}. The DNN hidden size is chosen from \{32, 64\} on the BookCrossing dataset and from \{64, 128, 256\} on the MovieLens-1M and MovieLens-25M datasets.
    \item \textbf{SASRec}~\cite{SASRec}. The number of attention heads is selected from \{1, 2, 4\}, and the number of attention layers is determined from \{1, 2, 3\}. The attention size is chosen from \{32, 64, 128\} on the BookCrossing dataset and from \{64, 128, 256\} on the MovieLens-1M and MovieLens-25M datasets. The number of DNN layers is picked from \{1, 2, 3\}, and the DNN hidden size is selected from \{32, 64\} on the BookCrossing dataset and from \{64, 128, 256\} on the MovieLens-1M and MovieLens-25M datasets.
    \item \textbf{DIN}~\cite{zhou2018deep}. The number of DIN attention layers and DNN layers are both selected from \{1, 2, 3\}. The DNN hidden size is chosen from \{32, 64\} on the BookCrossing dataset and from \{64, 128, 256\} on the MovieLens-1M and MovieLens-25M datasets.
    \item \textbf{SIM}~\cite{SIM}. The number of attention layers and DNN layers are both determined from \{1, 2, 3\}. The DNN hidden size is selected from \{32, 64\} on the BookCrossing dataset and from \{64, 128, 256\} on the MovieLens-1M and MovieLens-25M datasets.
\end{itemize}

\subsection{LM-based Models}
The structure of the pretrained language models is kept unchanged. And AdamW~\cite{adamw} optimizer is used for all the baselines. The detailed training settings are as follows:
\begin{itemize}[leftmargin=10pt]
   \item \textbf{CTR-BERT}~\cite{CTRBERT}. A two-tower architecture based on the BERT~\cite{devlin2018bert} model is utilized to encode user and item information separately. The training process involves 10 epochs. The batch size is fixed at 1024. The learning rate is set to $5\times 10^{-5}$, which decays linearly. The warmup ratio is 0.05.
    \item \textbf{P5}~\cite{P5} is a unified sequence-to-sequence framework that employs T5~\cite{T5} as the backbone pretrained language model for various recommendation tasks. In this study, P5 is applied solely for CTR prediction. The training consists of 10 epochs with a batch size of 32. The learning rate is chosen from $\{5 \times 10^{-4}, 1 \times 10^{-3}\}$ and decays linearly. The warmup ratio is 0.05. In line with the official P5 implementation, gradient clipping is performed with a threshold of 1.0.
    \item \textbf{PTab}~\cite{liu2022ptab} follows a pretrain-finetune approach based on the BERT~\cite{devlin2018bert} model. Initially, PTab further pretrains the BERT model using the masked language modeling objective on textualized CTR data. Subsequently, BERT is finetuned for CTR prediction as a text classification task. Consistent with the original paper, BERT is pretrained for 10 epochs with a batch size of 1024. The pretraining learning rate is $5\times 10^{-5}$, which decays linearly. The warmup ratio is 0.05. For finetuning, the process spans 10 epochs with a batch size of 1024. The finetuning learning rate starts at $5\times 10^{-5}$ and decays linearly. The warmup ratio is 0.01.

    \item \textbf{TALLRec}~\cite{bao2023tallrec} adopts the simple instruction tuning on pure textual prompts with LoRA. We use the same hyperparameter configurations for TALLRec and ReLLaX, \eg\ the learning rate, the LoRA rank and dropout rate. The prompt template is shown in Figure~\ref{fig: prompt simple}, where the user behavior sequence is constructed by the most recent $K$ items.

    \item \textbf{CoLLM}~\cite{zhang2023collm}. We keep the pretrained traditional CTR model and LoRA configurations for CoLLM the same as ReLLaX. Other training configurations, \eg\ learning rate and epoch, are also searched in the same range as ReLLaX.
    The prompt template is similar to ReLLaX, while CoLLM constructs the user behavior sequence by the most recent $K$ items.

    \item \textbf{iLoRA}~\cite{kong2024customizing}. Following the original paper~\cite{kong2024customizing}, we set the number of LoRA experts to 4. The other configurations, \eg\ the pretrained traditional CTR model, LoRA rank and training hyperparameters, are kept the same as ReLLaX. The prompt template is shown in Figure~\ref{fig: prompt simple}.

    \item \textbf{ReLLa}~\cite{lin2024rella}. We maintain the configurations of ReLLa used in the original paper~\cite{lin2024rella}, and we directly use the results of the original paper in this journal manuscript.

\end{itemize}